\documentclass[journal,10pt]{IEEEtran}

\IEEEoverridecommandlockouts
\usepackage{cite}
\usepackage{caption}
\usepackage{amsmath,amssymb,amsfonts}
\usepackage{algorithmic}
\usepackage{graphicx}
\usepackage{tabularx}
\usepackage{textcomp}
\usepackage{xcolor}
\usepackage{stfloats}
\usepackage{multirow}
\def\BibTeX{{\rm B\kern-.05em{\sc i\kern-.025em b}\kern-.08em
    T\kern-.1667em\lower.7ex\hbox{E}\kern-.125emX}}
\begin{document}

\title{A Novel 3D Non-Stationary GBSM for 6G THz Ultra-Massive MIMO Wireless Systems}

\author{Jun Wang, \textit{Student Member, IEEE}, Cheng-Xiang Wang, \textit{Fellow, IEEE}, Jie Huang, \textit{Member, IEEE}, Haiming Wang, \textit{Member, IEEE}, Xiqi Gao, \textit{Fellow, IEEE}, Xiaohu You, \textit{Fellow, IEEE}, and Yang Hao, \textit{Fellow, IEEE} 
\thanks{This work was supported by the National Key R\&D Program of China under Grant 2018YFB1801101, the National Natural Science Foundation of China (NSFC) under Grant 61960206006 and 61901109,  the Frontiers Science Center for Mobile Information Communication and Security, the High Level Innovation and Entrepreneurial Research Team Program in Jiangsu, the High Level Innovation and Entrepreneurial Talent Introduction Program in Jiangsu, the Research Fund of National Mobile Communications Research Laboratory, Southeast University, under Grant 2020B01 and 2021B02, the Fundamental Research Funds for the Central Universities under Grant 2242021R30001, and the EU H2020 RISE TESTBED2 project under Grant~872172. \textit{(Corresponding author: Cheng-Xiang Wang.)}}
\thanks{J. Wang, C.-X. Wang, J. Huang,  X. Gao, and X. You are with National Mobile Communications Research Laboratory, School of Information Science and Engineering, Southeast University, Nanjing, 210096, China, and also with the Purple Mountain Laboratories, Nanjing, 211111, China (email: {\{jun.wang, chxwang, j\_huang, xqgao, xhyu\}@seu.edu.cn}). }
\thanks{H. Wang is with State Key Laboratory of Millimeter Wave, School of Information Science and Engineering, Southeast University, Nanjing, 210096, China, and also with the Purple Mountain Laboratories, Nanjing, 211111, China (email: hmwang@seu.edu.cn).}
\thanks{Y. Hao is with the School of Electronic Engineering and Computer Science, Queen Mary University of London, London, E1 4NS, U.K. (e-mail: y.hao@qmul.ac.uk).}
}
\markboth{IEEE TRANSACTIONS ON VEHICULAR TECHNOLOGY,~Vol.~XX, No.~XX, MONTH~2021}%
{J. Wang \MakeLowercase{\textit{et al.}}: Bare Demo of IEEEtran.cls for IEEE Journals}

\maketitle

\begin{abstract}
Terahertz (THz) communication is now being considered as one of possible technologies for the sixth generation (6G) wireless communication systems. In this paper, a novel three-dimensional (3D) space-time-frequency non-stationary theoretical channel model is first proposed for 6G THz wireless communication systems employing ultra-massive multiple-input multiple-output (MIMO) technologies with long traveling paths. Considering frequency-dependent diffuse scattering, which is a special property of THz channels different from millimeter wave (mmWave) channels, the relative angles and delays of rays within one cluster will evolve in the frequency domain. Then, a corresponding simulation model is proposed with discrete angles calculated using the method of equal area (MEA). The statistical properties of the proposed theoretical and simulation models are derived and compared, showing good agreements. The accuracy and flexibility of the proposed simulation model are demonstrated by comparing the simulation results of the relative angle spread and root mean square (RMS) delay spread with corresponding measurements.
\end{abstract}

\begin{IEEEkeywords}
6G, THz GBSM, ultra-massive MIMO, long traveling paths, space-time-frequency non-stationarity 
\end{IEEEkeywords}

\section{Introduction}

The peak data rate and connection density of the fifth generation (5G) wireless communication system are 20 {gigabits per second (Gbps)} and 10$^6$ devices/km$^2$, respectively \cite{5Gsurvey}. It is expected that the sixth generation (6G) wireless system will reach {terabits per second (Tbps)} level in terms of the peak data rate and 10$^7$--10$^8$~devices/km$^2$ in terms of connection density \cite{J20_SCIS_XHYou_6G,WangCX1,WangCX2}. To meet the massive data flow and massive connectivity requirements in 6G, increasing transmission bandwidth and improving spectrum efficiency are potential solutions. Terahertz (THz) communication has the ability to provide more than one hundred {gigahertz (GHz)}  bandwidth\cite{RN532} and can theoretically achieve ultra-high transmission rate of 100 Gbps or even higher \cite{RN392}. By filling the gap between millimeter wave (mmWave) and optical wireless communication bands, THz communication has attracted great interests worldwide and been considered as one of the most promising 6G technologies. {
It has plenty of promising applications such as data center\cite{datacenter1,datacenter2} and kiosk downloading~\cite{RN462,RN467}, which support short-range~(about 1 m) communications between fixed devices and terminal equipment such as smart phones. THz communications can also be applied to smart rail systems\cite{Guanke1,Guanke2,Guanke3} by providing high data rate transmissions for passengers. In addition, the small size of THz devices makes nanoscale communications possible such as chip-to-chip\cite{RN599,RN143}, computer motherboard links\cite{motherboard}, and intra-body communications\cite{Haoyang1,Haoyang3,Haoyang4,Haoyang5}.}

THz channel models that can accurately reflect THz channel characteristics are prerequisite for the design, optimization, and performance evaluation of THz wireless communication systems. {The investigation of channel characteristics is very important for accurate channel modeling.} 
Due to the high frequency and large bandwidth of THz communications, propagation mechanisms of THz bands are quite different from those of lower frequency bands such as mmWave bands. The THz propagation mechanisms were studied in  \cite{RN308,RN356,RN197,RN364,RN346,RN394,RN369}. In \cite{RN308} and \cite{RN356}, THz measurements and modeling of multiple reflection effects in different materials were introduced. The reflection loss showed great dependence on frequency and materials and could be calculated by Kirchhoff theory. According to the measurement in \cite{RN197}, high-order paths were very hard to be detected due to high reflection loss which means that the number of multipaths is limited. {The diffusely scattered propagation was investigated in\cite{RN364,RN346,RN394,RN369,scattrer1}}. In \cite{RN364},  the detected signal powers in all directions for different materials were measured and simulated. Frequency-dependent scattering was also measured and simulated in \cite{RN346} and \cite{RN394}. The wavelength of THz waves is comparable to the roughness of some common materials. As a result, more power is diffusely scattered when frequency increases. {Most of the diffusely scattered rays surround the specular reflected paths. In this case, scattered rays far away from the specular reflection points can be neglected\cite{RN369}}.  


{Apart from investigations on THz channel characteristics, a number of THz channel models were proposed for THz communication systems.} In~\cite{RN207}, a  multi-ray ray tracing channel model for THz indoor communication was presented and 
validated with experiments. In \cite{RN229}, a three-dimensional (3D) time-variant THz channel model based on ray tracing was investigated for dynamic environments. However, channel models based on ray tracing methods are not general and flexible. In \cite{RN194}, a stochastic THz indoor channel model considering the frequency dispersion was presented and verified by ray tracing. In \cite{RN195}, the authors investigated root mean square (RMS) delay spread and angular spread that were modeled by second order polynomial parameters for THz indoor communications. 
However, the existing stochastic THz channel models cannot show the unique propagation characteristics of THz bands such as frequency-dependent scattering.

{In the existing THz channel models, massive multiple-input multiple-output (MIMO) {were} rarely mentioned. However,  in THz communication systems, ultra-massive MIMO technologies employing hundreds or even thousands of antennas are considered as one of the solutions to compensate the high path loss \cite{molisch1} and are expected to be utilized in 6G\cite{WangCX1,WangCX2}}. 
Massive MIMO channel models and measurements considering spatial non-stationarity were studied in \cite{RN321,massiveMIMO_bianji,RN322,massiveMIMO1,massiveMIMO2,massiveMIMO3}.  
In~\cite{RN321}, massive MIMO channel measurements and models were summarized. In \cite{RN161,HuangjieJSAC,HuangjieCSCI}, channel measurements for massive MIMO channels in different scenarios at mmWave were presented. A novel 3D {geometric-based stochastic model (GBSM)} based on homogeneous Poisson point process was proposed for mmWave channels\cite{HuangjieCSCI}. A GBSM for mobile-to-mobile scenarios for mmWave bands was presented in \cite{HeRuisiTVT2018}. However, the existing massive MIMO channel models have only considered characteristics of sub-6~GHz and mmWave bands, and are not suitable for THz communication systems because propagation mechanisms in THz bands are quite different from those of lower frequencies. 

In addition, time variant channels caused by long traveling paths also need to be characterized in THz communication systems. {It should be noted that for a realistic application of THz communications, the range of movement is not necessarily very long. For example, in some typical application scenarios such as movement of  human body,  even a short traveling path of several meters is quite long compared with the wavelength of THz signals.} Unlike high mobility scenarios such as high-speed train and vehicle-to-vehicle communications in fifth generation (5G) systems, THz communications are usually applied in relatively static environments. {The non-stationarity in the time domain is caused by the continuous moving of the transmitter or receiver over a long distance.}

To the best of the authors’ knowledge, stochastic space-time-frequency non-stationary GBSMs for massive MIMO THz channels are still missing in the literature. We have done some early investigations in \cite{WCNC20_WJ,JSAC_WJ}. However, the simulation models of THz channels were not studied in \cite{JSAC_WJ}. The major contributions and novelties of this paper are listed as follows. 

\begin{enumerate}
	\item A novel theoretical space-time-frequency non-stationary GBSM for ultra-massive MIMO THz channels considering long traveling paths is first proposed. The non-stationarity in space, time, and frequency domains caused by ultra-massive MIMO, long traveling paths, and large bandwidths, respectively, are considered. 
	\item The statistical properties of the theoretical model such as space-time-frequency correlation function (STFCF), time autocorrelation function (ACF), spatial cross-correlation function (CCF), and frequency correlation function (FCF) are derived.
	\item The corresponding simulation model with discrete angles generated by the method of equal area (MEA) is then proposed. 
	Its statistical properties are derived, verified by simulation results, and compared with those of the theoretical model, showing good agreements. Some simulated statistical properties are also compared with corresponding measurements, illustrating the validity and flexibility of the proposed THz simulation model. 
\end{enumerate}

The remainder of this paper is organized as follows. 
In Section~\uppercase\expandafter{\romannumeral2}, the proposed theoretical THz GBSM is described in detail and the channel impulse response (CIR) is presented. In Section~\uppercase\expandafter{\romannumeral3}, statistical properties of the theoretical Thz channel model are derived. Then, the corresponding simulation model is proposed with discrete angles calculated using the MEA and its statistical properties are derived in Section~\uppercase\expandafter{\romannumeral4}. In Section~\uppercase\expandafter{\romannumeral5}, statistical properties of the theoretical model and simulation model, simulation results, and measurements are compared and discussed. Finally, conclusions are drawn in Section~\uppercase\expandafter{\romannumeral6}.


\section{A 3D THz Non-Stationary Theoretical Model}

\subsection{Description of the  Channel Model}

\begin{figure*}[tb]
	\centerline{\includegraphics[width=1\textwidth]{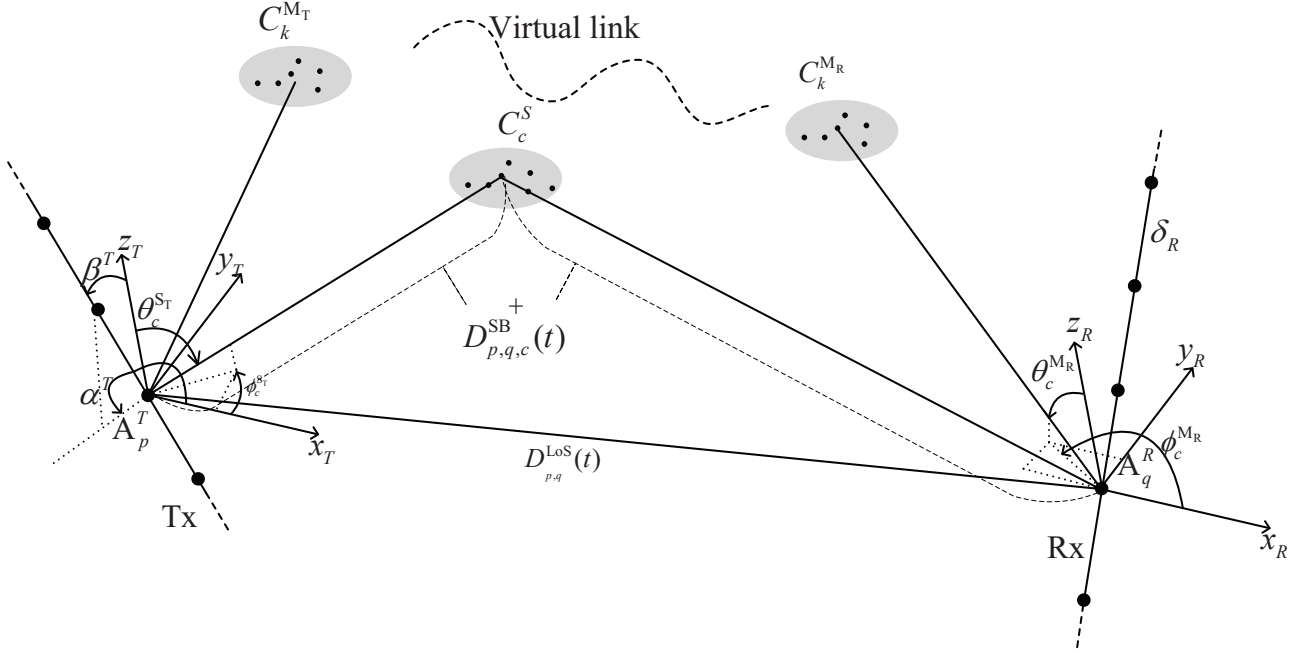}} 
	\caption{A 3D THz GBSM for massive MIMO communication systems.}
	\label{fig_1}
\end{figure*}

In this sub-section, a 3D non-stationary THz massive MIMO channel model is proposed. The proposed  model is illustrated in Fig. \ref{fig_1}. Both transmitted antennas (Tx) and received antennas (Rx) are equipped with a massive number of antenna elements. To simplify the calculation, uniform linear arrays (ULAs) are assumed in this model. We assume that Tx has $N_\text{T}$ antenna elements with the same distance $\delta_\text{T}$. It should be noted that only propagation model is studied so that each element in the array is assumed omni-directional. The direction of the Tx is represented by the elevation and azimuth angles $\beta^\text{T}$ and $\alpha^\text{T}$, respectively. Rx is represented similarly.  The $p$th transmit and $q$th receive antenna element are denoted by  $A_p^\text{T}$ and $A^\text{R}_q$, respectively. The position vector of the  $A_p^\text{T}$ from the center of the transmitted array  $\overrightarrow{\delta}_p $ and  $A_q^\text{R}$ from the center of the received array  $\overrightarrow{\delta}_q $ are expressed by

\begin{equation}
\overrightarrow{\delta}_p={\delta}_p\cdot\left[ 
\begin{matrix}
\cos\beta^\text{T}\cos\alpha^\text{T}\\
\cos\beta^\text{T}\sin\alpha^\text{T}\\
\sin\beta^\text{T}
\end{matrix}
\right]
\end{equation}\begin{equation}
\overrightarrow{\delta}_q=\delta_q\cdot\left[ 
\begin{matrix}
\cos\beta^\text{R}\cos\alpha^\text{R}\\
\cos\beta^\text{R}\sin\alpha^\text{R}\\
\sin\beta^\text{R}
\end{matrix}
\right]
\end{equation}
with ${\delta}_p=\frac{N_T-2p+1}{2}\delta_\text{T}$ and $\delta_q=\frac{N_R-2q+1}{2}\delta_\text{R}$. 
In addition, the speed of Rx is denoted by $\text{v}^\text{R}$.    The elevation and azimuth angles of the speed are represented by $\zeta^\text{R}$ and $\xi^\text{R}$, respectively.

The concept of cluster is proposed\cite{RN248} to simplify the analysis of multipaths. In THz band, the wavelength of the carrier frequency is less than one millimeter and comparable to the surface  roughness of some common objects such as furniture and walls. In this model, each cluster is comprised of diffusely scattering rays from  the rough surface.  The center of cluster is considered  as the specular reflected point. According to the study in\cite{RN364}, the power and angle of these clusters are frequency-dependent. 

The received signal at the Rx is composed of of the line of sight (LoS),  $C_\text{S}$ single-bounce clusters (SBCs), and $C_\text{M}$ multi-bounce clusters (MBCs). 
In the picture, only the $c$th SBC and $k$th MBC are represented for clarity. Each MBC consists of two clusters at both Tx and Rx connected by a virtual link. For the $k$th MBC the transmit-side and receive-side MBC are denoted as $C^{\text{M}_\text{T}}_k$ and  $C^{\text{M}_\text{R}}_k$, respectively. The $c$th SBC is denoted by $C^\text{S}_c$. Both SBCs and MBCs are comprised of infinite scatterers. The main parameters are defined in Table~\ref{tab1}.

Considering the frequency domain non-stationarity in the propagation channel, the whole band is divided into $N_F$ small sub-bands in which the channel is considered stationary in frequency domain\cite{RN207}. 
The CIR of sub-band  $h_{p,q}(t,\tau,f_i)$ is represented as
\begin{equation}\label{eq_CIR}
\begin{split}
\hspace{3mm}
h_{p,q}(t,\tau,f_{i})=&h_{p,q}^{\text{LoS}}(t,\tau,f_{i})+
\sum_{c=1}^{C_\text{SB}}h_{p,q,c}^{\text{SB}}(t,\tau,f_{i})\\&+\sum_{k=1}^{C_\text{MB}}h_{p,q,k}^{\text{MB}}(t,\tau,f_{i})
\end{split}
\end{equation}
where LoS, SB, and MB refer to line of sight, single bounce, and multi bounce components, respectively.
\renewcommand{\arraystretch}{1.2}
\begin{table*}[t]
	\centering
	\setlength{\belowcaptionskip}{0.2cm}
	\caption{Definitions of main parameters for the proposed THz massive MIMO channel model.}
	\begin{tabular}{|l|l|}			
		\hline
		\textbf{Parameters}&\textbf{Definitions}\\
		\hline
		$\delta_\text{T}$,$\delta_\text{R}$&Distances between consecutive elements in the Tx ans Rx array, respectively
		\\
		\hline
		$D$& Distance between the center of $\text{Tx}$ and  $\text{Rx}$\\
		\hline
		$D_{p,q}^\text{LoS}(t)$&Distance from ${A}_p^\text{T}$ to  ${A}_q^\text{R}$
		\\\hline
		$D_{p,q,c}^\text{SB}(t)$&Total distance of the path: ${A}_p^\text{T}$-$C_c^\text{S}$-${A}_q^\text{R}$
		\\\hline
		$D_{p,q,c,l}^\text{SB}(t,f_i)$&Total distance of the path: ${A}_p^\text{T}$-$C_{c,l}^\text{S}$-${A}_q^\text{R}$
		\\\hline
		$D_{p,q,k}^\text{MB}(t)$& Total distance of the path: ${A}_p^\text{T}$-$C_k^{\text{M}_\text{T}}$-$C_k^{\text{M}_\text{R}}$-${A}_q^\text{R}$
		\\\hline
		$D_{p,q,k,m}^\text{MB}(t)$& Total distance of the path: ${A}_p^\text{T}$-$C_{k,m}^{\text{M}_\text{T}}$-$C_{k,m}^{\text{M}_\text{R}}$-${A}_q^\text{R}$
		\\\hline
		$\beta^{\text{T}(\text{R})}$, $\alpha^{\text{T}(\text{R})}$&Elevation and azimuth angles of the Tx(Rx) array, respectively
		\\\hline
		$\theta^\text{LoS}$, $\phi^\text{LoS}$&Elevation and azimuth angles of the Rx with respect to the Tx, respectively
		\\\hline
		$\theta^{\text{S}_\text{T(R)}}_{p(q),c}(t)$, $\phi_{p(q),c}^{\text{S}_\text{T(R)}}(t)$&Elevation and azimuth angles of $C_c^{\text{S}_\text{T(R)}}$, respectively
		\\\hline
		$\theta^{\text{S}_\text{T(R)}}_{p(q),c,l}(t,f_i)$, $\phi^{\text{S}_\text{T(R)}}_{p(q),c,l}(t,f_i)$&Elevation and azimuth angles of the scatterer $C_{c,l}^{\text{S}_\text{T(R)}}$, respectively
		\\\hline
		$\theta^{\text{M}_\text{T(R)}}_{p(q),k}(t)$, $\phi^{\text{M}_\text{T(R)}}_{p(q),k}(t)$&Elevation and azimuth angles of $C_k^{\text{M}_\text{T(R)}}$, respectively
		\\\hline
		$\theta^{\text{M}_\text{T(R)}}_{p(q),k,m}(t,f_i)$, $\phi^{\text{M}_\text{T(R)}}_{p(q),k,m}(t,f_i)$&Elevation and azimuth angles of the scatterer $C_{k,m}^{\text{S}_\text{T(R)}}$, respectively
		\\\hline
		
		$\zeta^\text{R}, \xi^\text{R}$&Elevation and azimuth angles of the velocity of the Rx, respectively
		\\\hline
		
	\end{tabular}
	
	\label{tab1}
\end{table*}
In (\ref{eq_CIR}), the LoS component is represented as
\begin{equation}\label{eq_LoS}
\begin{split}
h_{p,q}^{\text{LoS}}(t,\tau,f_{i})=&\sqrt{\frac{K}{K+1}}\times e^{-j(2\pi f_{i}\tau_{p,q}^{\text{LoS}}(t))}\delta(\tau-\tau_{p,q}^{\text{LoS}}(t))
\end{split}
\end{equation}
where $K$ is the  Rician factor. The delay $\tau_{p,q}^{\text{LoS}}(t)$ is given by  $D_{p,q}^{\text{LoS}}(t)/c_0$ where $c_0$ refers to  the speed of light. The distance between ${A}_p^\text{T}$ and  ${A}_q^\text{R}$ is calculated by the vector $D_{p,q}^{\text{LoS}}(t)=\left \|\overrightarrow{\textbf{D}}_{p,q}^{\text{LoS}}(t)\right \|$ where $\left \|\cdot\right \|$ calculates the Frobenius norm. The $\overrightarrow{\textbf{D}}_{p,q}^{\text{LoS}}(t)$ is expressed as 
\begin{equation}\label{D_LoS}
\begin{split}
\overrightarrow{\textbf{D}}_{p,q}^{\text{LoS}}(t) = \overrightarrow{\textbf{D}}+\overrightarrow{\delta}_q-\overrightarrow{\delta}_p+\overrightarrow{\textbf{v}^\text{R}}\cdot t
\end{split}
\end{equation}
where
\begin{equation}
\overrightarrow{\textbf{D}}=D\cdot\left[ 
\begin{matrix}
\cos\theta^\text{LoS}\cos\phi^\text{LoS}\\
\cos\theta^\text{LoS}\sin\phi^\text{LoS}\\
\sin\theta^\text{LoS}
\end{matrix}
\right]
\end{equation}
\begin{equation}
\overrightarrow{\textbf{v}^\text{R}}=\text{v}^\text{R}\left[ 
\begin{matrix}
\cos\zeta^\text{R}\cos\xi^\text{R}\\
\cos\zeta^\text{R}\sin\xi^\text{R}\\
\sin\zeta^\text{R}
\end{matrix}
\right].
\end{equation}

The SB components and the MB components can be expressed as
\begin{equation}\label{eq_SB}
\begin{split}
h_{p,q,c}^{\text{SB}}(t,\tau,f_{i})=&\sqrt{\frac{1}{K+1}}\ \lim_{L_c\rightarrow\infty}\sum_{l=1}^{L_c} \sqrt{\frac{P^\text{SB}_c}{L_c}}\\&e^{-j(2\pi f_{i}\tau_{p,q,c,l}^{\text{SB}}(t)-\Theta^\text{SB}_{c,l})}\delta(\tau-\tau_{p,q,c,l}^{\text{SB}}(t))
\end{split}
\end{equation}
\begin{equation}\label{eq_MB}
\begin{split}
h_{p,q,k}^{\text{MB}}(t,\tau,f_{i})&=\sqrt{\frac{1}{K+1}}\lim_{M_k\rightarrow\infty}\sum_{m=1}^{M_k} \sqrt{\frac{P^\text{MB}_k}{M_k}}\\&e^{-j(2\pi f_{i}\tau_{p,q,k,m}^{\text{MB}}(t)-\Theta^\text{MB}_{k,m})}\delta(\tau-\tau_{p,q,k,m}^{\text{MB}}(t))
\end{split}
\end{equation}
where $\Theta^\text{SB}_{c,l}$ and $\Theta^\text{MB}_{k,m}$ are 
uniformly distributed over $(0,2\pi]$. 
{The $L_c$ and $M_k$ refer to the   number of rays for the $c$th single-bounce cluster and the $k$th multi-bounce cluster, respectively. In addition,} $P^\text{SB}_c$ and $P^\text{MB}_k$ are the frequency-dependent power of corresponding clusters. The symbols $\tau_{p,q,c,l}^{\text{SB}}(t)$ and $\tau_{p,q,k,m}^{\text{MB}}(t)$ are the delays of corresponding rays.

The delay $\tau_{p,q,c,l}^{\text{SB}}(t)$ of the $l$th ray in the $c$th SBC is composed of two parts, i.e.,
\begin{equation}
\tau_{p,q,c,l}^{\text{SB}}(t)=\tau_{p,q,c}^{\text{SB}}(t)+\Delta\tau_{p,q,c,l}^{\text{SB}}(t)
\end{equation}
where $\tau_{p,q,c}^{\text{SB}}(t)$ is the delay of the path ${A}_p^\text{T}$-$C_c^\text{S}$-${A}_q^\text{R}$. Here, $C_c^\text{S}$ represents the center of the cluster so this path is the specular reflection path of this cluster. The relative delay with respect to this  specular reflection path is denoted by $\Delta\tau_{p,q,c,l}^{\text{SB}}(t)$. 

To obtain the $\Delta\tau_{p,q,c}^{\text{SB}}(t)$, we need to generate the initial delay between the center of Tx and Rx arrays via all the clusters at initial time $\tau_{c}^{\text{SB}}$.
The initial cluster delay $\tau_{c}^{\text{SB}}(t_0)$ is generated by random variables $\Delta \tau_{c,\text{SB}}$\cite{RN194}, where $\Delta \tau_{c,\text{SB}}$  is defined as the time interval of arrival between two adjacent clusters for the first order cluster and the second order cluster, respectively. 
For the first cluster, $\Delta \tau_{1,\text{SB}}$ is the time interval compared to the LoS path. So, we have
\begin{equation}
\tau_{c}^\text{SB}(t_0) = \begin{cases}
D/c_0+\Delta \tau_{c,\text{SB}},&c=1;\\
\tau_{c-1}+\Delta \tau_{c,\text{SB}},&2\leq c\leq C_\text{SB}.
\end{cases}
\end{equation}
A similar method is used to generate MBCs expressed as 
\begin{equation}
\tau_{k}^\text{MB}(t_0) = \begin{cases}
D/c_0+\Delta \tau_{k,\text{MB}},&k=1;\\
\tau_{k-1}+\Delta \tau_{k,\text{MB}},&2\leq k\leq C_\text{MB}.
\end{cases}
\end{equation}
Here, $\Delta \tau_{c,\text{SB}}$ and $\Delta \tau_{k,\text{MB}}$ are  exponential random variables with the mean values $\mu_{\Delta \tau_{\text{SB}}}$ and $\mu_{\Delta \tau_{\text{MB}}}$, respectively. 
\begin{figure}[t]
	\centerline{\includegraphics[width=0.5\textwidth]{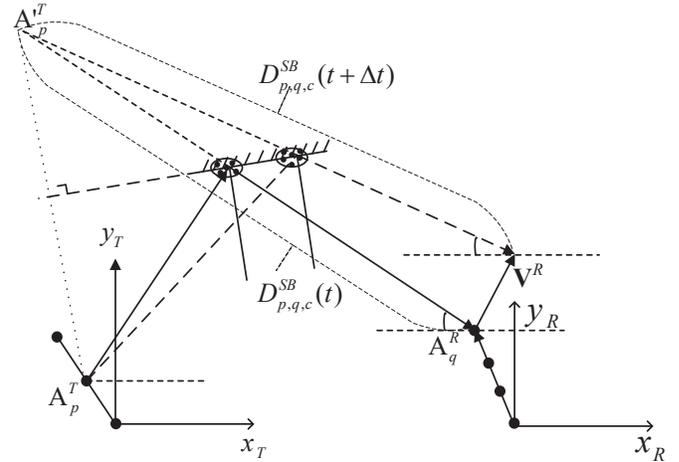}}
	\caption{Geometric relationship of the time-variant path via the SBC.}
	\label{fig_2}
\end{figure}

Considering the long traveling paths, after we generate the initial delay for SBC at $t$, the total delay of the cluster at $t+\Delta t$ can be calculated according to the geometric relationship in Fig. \ref{fig_2}. {When the Tx is fixed and the Rx moves, we extend the reflection surface closest to the Tx and obtain the mirror point of the Tx. When the Rx moves in a short period, the mirror point keeps static and the cluster slides for a short length at the surface. The solid line is the initial path and the dashed line represents the path after the Rx moves. Similarly, when the Rx is fixed and the Tx moves, we can also find the mirror point of the Rx for calculation. This method jointly considers the movement of the clusters and the Tx/Rx.} 

{We denote ${A'}_p^\text{T}$ to represent the mirror point of $A_p^\text{T}$.} It is clear that the distance from ${A'}_p^\text{T}$ to $A_q^\text{R}$ at time $t$ equals $D^\text{SB}_{p,q,c}(t)$. The vector $\overrightarrow{{D'}}^\text{SB}_{p,q,c}(t)$ from ${A'}_p^\text{T}$ to $A_q^\text{R}$ is 
\begin{equation}
\overrightarrow{{D'}}^\text{SB}_{p,q,c}(t) =D^\text{SB}_{p,q,c}(t)\cdot \left[ 
\begin{matrix}
\cos\theta_{q,c}^{\text{S}_\text{R}}\cos\phi_{q,c}^{\text{S}_\text{R}}\\
\cos\theta_{q,c}^{\text{S}_\text{R}}\sin\phi_{q,c}^{\text{S}_\text{R}}\\
\sin\theta_{q,c}^{\text{S}_\text{R}}
\end{matrix}
\right]. 
\end{equation}
The vector $\overrightarrow{{D'}}^\text{SB}_{p,q,c}(t+\Delta t)$ at time $t+\Delta t$  is easily calculated as
\begin{equation}
\begin{split}
&\overrightarrow{{D'}}^\text{SB}_{p,q,c}(t+\Delta t)=\overrightarrow{{D'}}^\text{SB}_{p,q,c}(t)+\overrightarrow{\textbf{v}^\text{R}} \Delta t\\&=D^\text{SB}_{p,q,c}(t) \left[ 
\begin{matrix}
\cos\theta_{q,c}^{\text{S}_\text{R}}\cos\phi_{q,c}^{\text{S}_\text{R}}\\
\cos\theta_{q,c}^{\text{S}_\text{R}}\sin\phi_{q,c}^{\text{S}_\text{R}}\\
\sin\theta_{q,c}^{\text{S}_\text{R}}
\end{matrix}
\right]+v_R\Delta t\cdot\left[ 
\begin{matrix}
\cos\zeta^\text{R}\cos\xi^\text{R}\\
\cos\zeta^\text{R}\sin\xi^\text{R}\\
\sin\zeta^\text{R}
\end{matrix}
\right].
\end{split}
\end{equation}
The new distance from ${A'}_p^\text{T}$ to $A_q^\text{R}$ at time $(t+\Delta t)$ denoted by $D^\text{SB}_{p,q,c}(t+\Delta t)$  can be calculated by 
\begin{equation}
\begin{split}
D^\text{SB}_{p,q,c}(t+\Delta t) = \left \|\overrightarrow{{D'}}^\text{SB}_{p,q,c}(t+\Delta t)\right \|.
\end{split}
\end{equation}
{For the MBCs, when the Rx is moving, the mirror point of the Tx can be obtained by extending the line between  Rx and last reflection point with the total path length. The virtual link is contained in the extended line so that this method can also be adopted in the MBCs}

Considering the spatial non-stationarity of ultra-massive MIMO, the distance from ${A'}_p^\text{T}$ to $A_q^\text{R}$ can be analogously obtained by using  $\overrightarrow{\delta}_q $ to replace $\overrightarrow{\textbf{v}^\text{R}} \Delta t$ and can be calculated by  
\begin{equation}
\begin{split}
&{{D}}^\text{SB}_{p,q,c}(t)=\left \|\overrightarrow{{D'}}^\text{SB}_{c}(t)+\overrightarrow{{\delta}_p}\right \|
\end{split}
\end{equation}
where

\begin{equation}
\begin{split}
\overrightarrow{{D'}}^\text{SB}_{c}(t) = 
D^\text{SB}_{c}(t) \left[ 
\begin{matrix}
\cos\theta_{c}^{\text{S}_\text{R}}\cos\phi_{c}^{\text{S}_\text{R}}\\
\cos\theta_{c}^{\text{S}_\text{R}}\sin\phi_{c}^{\text{S}_\text{R}}\\
\sin\theta_{c}^{\text{S}_\text{R}}
\end{matrix}
\right].
\end{split}
\end{equation}
Here $D^\text{SB}_{c}(t)$ {represents} the total distance from the center of Tx to Rx via $C_c^\text{S}$.

The power of the cluster  $P_{c}^\text{SB}$ and $P_{k}^\text{MB}$ is generated according to the delay of the paths\cite{RN194}. We use SBC as an example and they can be expressed as 
\begin{equation}
\label{power_cluster}
\begin{split}
P_{c}^\text{SB}(\text{dB}) =-n_\tau\cdot(\tau_{c}^\text{SB}-\tau^\text{LoS})+\Delta a_i
\end{split}
\end{equation} 
where $n_\tau$ is the temporal decay coefficient. Moreover,  $\Delta a_i$ is a Gaussian distributed random variable describing the random deviation of different clusters. After we generate the power of all clusters, we need to normalize these  by the total~power. 



\subsection{Frequency-Dependent Parameters}

\begin{figure}[t]
	\centerline{\includegraphics[width=0.5\textwidth]{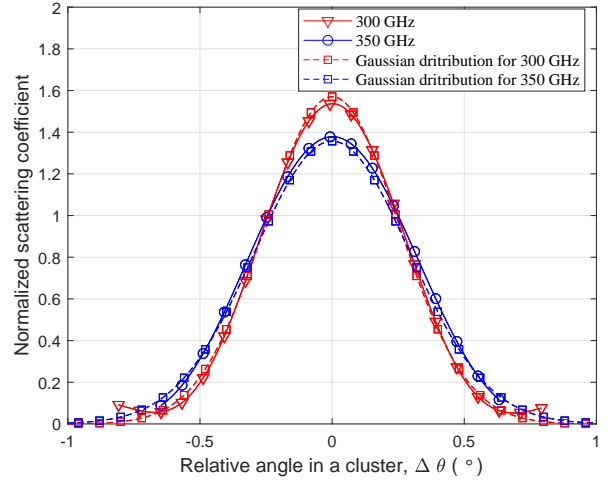}}
	\caption{The comparison of normalized scattering coefficients at different frequencies and fitting with Gaussian distribution ($\sigma$ = 0.25 at 300 GHz, $\sigma$ = 0.29 at 350 GHz, $l_{\text{corr}} = 2.3$  mm, $l_x = l_y = 10 l_{\text{corr}}$, $\sigma_h$ = 0.13 mm).}
	\label{Scattering_Gaussian}
\end{figure}

{In THz bands, wavelength is comparable to the roughness of the reflection surface.  In addition, scattering rays are closely related to the frequency, which is quite different from lower frequencies. The distribution of reflected power from one incident wave can be measured by scattering coefficient which was investigated in \cite{RN369}. For an incident wave, {the path with the strongest reflection} is the specular reflection. However, the scattering around the specular reflection path also need to be considered. } 
In~\cite{RN369}, the average scattering coefficient  $\left\langle \rho\rho^*\right\rangle_\infty $ is given, and can be calculated as
\begin{equation}
\left\langle \rho\rho^*\right\rangle_\infty=e^{-g}\cdot\left( \rho_0^2+\frac{\pi l_{\text{corr}}^2F^2}{A}\sum_{m=1}^{\infty}\frac{g^m}{m!m}e^{-\frac{v_{xy}^2l_{\text{corr}}^2}{4m}}\right)
\end{equation}
with
\begin{equation}
\rho_0=sinc\left(v_xl_x \right)\cdot sinc\left(v_yl_y \right) 
\end{equation}
\begin{equation}
v_x=k\cdot\left(\sin(\theta_1)-\sin (\theta_2)\cos (\theta_3) \right) 
\end{equation}
\begin{equation}
v_y=k\cdot\left(-\sin(\theta_2)\sin(\theta_3)\right) 
\end{equation}
\begin{equation}
v_{xy}=\sqrt{v_x^2+v_y^2}
\end{equation}
\begin{equation}
F=\dfrac{1+\cos(\theta_1)\cos(\theta_2)-\sin(\theta_1)\sin(\theta_2)\cos(\theta_3)}{\cos(\theta_1)(\cos(\theta_1)+\cos(\theta_2))} 
\end{equation}
\begin{equation}
g=k^2\sigma_h^2(\cos(\theta_1)+\cos(\theta_2))^2.
\end{equation}
Here, $l_x \cdot l_y$ calculates the the rectangular surface area. The parameters $\sigma_h$ and $l_{\text{corr}}$ represent the standard deviation of heights and the surface correlation length, respectively, given in \cite{RN369}. $\theta_1$, $\theta_2$, and $\theta_3$ {represent} the incident angle, exit angle, and the angle between incident plane and exit plane, respectively.  Parameter $k$ is the wave number  decided by frequency.  According to these equations, 
the normalized scattering coefficients  at different frequencies are compared in Fig.~\ref{Scattering_Gaussian}. It should be noticed that only the main lobe of the scattering beam is taken into consideration. From Fig.~\ref{Scattering_Gaussian}, we find that Gaussian distribution fits well with the main lobe at different frequencies. The standard deviation $\sigma$ of scattered wave at 300 GHz is 0.25 and is 0.29 at 350 GHz. 
This means that the relative angle in each cluster can be considered as zero mean Gaussian distributed random variables. 

Here,  $\phi_{p,c,l}^{\text{S}_\text{T}}$ is shown  as an example. The azimuth angle of the $l$th ray in the $c$th cluster is 
\begin{equation}
\phi_{p,c,l}^{\text{S}_\text{T}}=\phi_{p,c}^{\text{S}_\text{T}}+\Delta\phi_{p,c,l}^{\text{S}_\text{T}}
\end{equation}
where $\Delta\phi_{p,c,l}^{\text{S}_\text{T}}$ is the relative angle for the $l$th ray in the $c$th cluster with respect to $\phi_{p,c}^{\text{S}_\text{T}}$ and follows a Gaussian distribution with standard deviation  of $\sigma_{c,\phi}^{\text{S}_\text{T}}(f_i)$. 
The standard deviation $\sigma_{c,\phi}^{\text{S}_\text{T}}(f_i)$ is considered as an exponential distributed random variable according to  \cite{RN195}
\begin{equation}
f_{\mu}(x)=\frac{1}{\mu}e^{-\frac{x}{\mu}}
\end{equation}
where $\mu$ is the mean value of azimuth for the $c$th cluster described by parameter  $\mu_{c,\phi}^{\text{S}_\text{T}}$. 
Relative angles need to be regenerated at new frequencies and can be modeled as 
\begin{equation}
\sigma_{c,\phi}^{\text{S}_\text{T}}(f_{i})=\sigma_{c,\phi}^{\text{S}_\text{T}}(f_{0})\times(f_i/f_0)^{\rho_{\phi}^{\text{S}_\text{T}}}
\end{equation}
where $\rho_{\phi}^{\text{S}_\text{T}}$ represents the frequency-dependent factor for relative angle. The relative  elevation and azimuth angles of clusters are considered as independent.

According to the relative angle, the total time of arrival of different scattering paths can be calculated by the geometric  relationship as
\begin{equation}
	D^{\text{SB}}_{p,q,c,l}=\sqrt{{D^{\text{SB,V}}_{p,q,c,l}}^2+{D^{\text{SB,H}}_{p,q,c,l}}^2}
	\label{29}
\end{equation}
where ${D^{\text{SB,V}}_{p,q,c,l}}$ and ${D^{\text{SB,H}}_{p,q,c,l}}$ represent the vertical and  horizontal distances of the path length, respectively. They can be calculated by
\begin{equation}
	\label{30}
\begin{split}
{D^{\text{SB,V}}_{p,q,c,l}}= &{D^{\text{SB}}_{p,q,c}} \sin(\theta^{\text{S}_\text{R}}_{q,c})r^{\text{S}_\text{R}}_{c}/\cos(\Delta\theta_{p,c,l}^{\text{S}_\text{R}})
\\+&{D^{\text{SB}}_{p,q,c}} \sin(\theta^{\text{S}_\text{R}}_{q,c})r^{\text{S}_\text{T}}_{c}/\cos(\Delta\theta_{p,c,l}^{\text{S}_\text{T}})\end{split}
\end{equation}
\begin{equation}
	\label{31}
\begin{split}
{D^{\text{SB,H}}_{p,q,c,l}}= &{D^{\text{SB}}_{p,q,c}} \cos(\theta^{\text{S}_\text{R}}_{q,c})r^{\text{S}_\text{R}}_{c}/\cos(\Delta\phi_{p,c,l}^{\text{S}_\text{R}})
\\+&{D^{\text{SB}}_{p,q,c}} \cos(\theta^{\text{S}_\text{R}}_{q,c})r^{\text{S}_\text{T}}_{c}/\cos(\Delta\phi_{p,c,l}^{\text{S}_\text{T}})\end{split}
\end{equation}
where $r^{\text{S}_\text{T(R)}}_{c}$ is the ratio of the distance between the cluster and the Tx(Rx) to the  total distance. For SBCs, it is clear that $r^{\text{S}_\text{T}}_{c}$+$r^{\text{S}_\text{R}}_{c}$ = 1. For MBCs,  $r^{\text{S}_\text{T}}_{c}$+$r^{\text{S}_\text{R}}_{c}\textless1$.
{When relative angles are updated in the frequency domain, corresponding total distances are also updated according to the geometric relationship in (\ref{29})-(\ref{31}).}

%


\subsection{Channel Transfer Function}

\begin{figure}[t]
	\centerline{\includegraphics[width=0.45\textwidth]{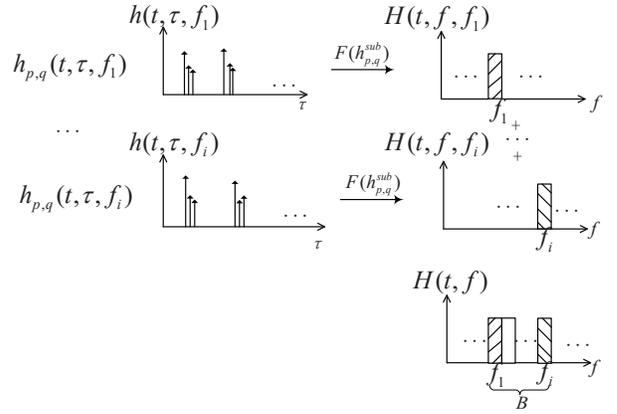}}
	\caption{An example of adding up two sub-band channels.}
	\label{CTF}
\end{figure}

In this model, we divide the whole band into many small sub-bands with different CIR. As a result, we need to calculate the CTF of each sub-band and then combine them together. Fig.~\ref{CTF} is an example of adding up two sub-band channels. Similarly, we can add up more sub-band channels in the same manner. 

The CTF of sub-band channel $H_{p,q}(t,f,f_{i})$ is calculated as the Fourier transformation of the sub-band CIR, i.e.,
\begin{equation}
{H}_{p,q}(t,f,f_{i})= \int_{-\infty}^{\infty}{h}_{p,q}(t,\tau,f_{i})e^{-j2\pi f\tau}d\tau,  f\in f_{i}.
\end{equation} 
The CTF of the whole band is given by
\begin{equation}
{H}_{p,q}(t,f) = \sum_{i=1}^{N_F}{{H}_{p,q}(t,f,f_{i})}.
\end{equation}

\section{Statistical Properties of the Theoretical Model}
\subsection{STFCF}
{The correlation between two arbitrary CIRs,  $h_{p,q}(t,\tau,f_i)$ and $h_{p',q'}(t+\Delta t,\tau,f_{i}+\Delta f)$ at different times, spaces, and frequencies can be calculated by superimposing the correlation of all the clusters with the assumption of uncorrelated scattering.} The correlation function of two arbitrary channels can be calculated as follows
\begin{equation}\label{STFCF}
\begin{split}
&R_h(p,q,p',q',t,\Delta t,f_i, \Delta f) =\\ &E\left[h_{p,q}(t,\tau,f_i)h^*_{p',q'}(t+\Delta t,\tau,f_{i}+\Delta f) \right] 
\end{split}
\end{equation}
where $(\cdot)^*$ is the complex conjugate operation, and {$E[\cdot]$ calculates the expectation value by taking all 3D directions at both Tx and Rx sides into consideration}. With the assumption of uncorrelated scatterers, the STFCF can be also written as
\begin{equation}
\begin{split}
&R_h(p,q,p',q',t,\Delta t,f_i, \Delta f) =R_h^\text{LoS}(p,q,p',q',t,\Delta t,f_i, \Delta f)\\&+\sum_{c=1}^{C_\text{SB}}R_{h,c}^\text{SB}(p,q,p',q',t,\Delta t,f_i, \Delta f) \\&+\sum_{k=1}^{C_\text{MB}}R_{h,k}^\text{MB}(p,q,p',q',t,\Delta t,f_i, \Delta f).
\end{split}
\end{equation}

--In the LoS case,
\begin{equation}
\begin{split}
&R_h^\text{LoS}(p,q,p',q',t,\Delta t,f_i, \Delta f)=\\&\frac{K}{K+1}e^{j2\pi(f_i\tau_{p,q}(t)-(f_i+\Delta f)\tau_{p',q'}(t+\Delta t))}.
\end{split}
\end{equation}

--In the SB case,
\begin{equation}
\begin{split}
&R_{h,c}^\text{SB}(p,q,p',q',t,\Delta t,f_i, \Delta f)=\frac{P_c^\text{SB}}{K+1}\\&\int_{-\pi}^{\pi}\int_{-\pi/2}^{\pi/2}\int_{-\pi}^{\pi}\int_{-\pi/2}^{\pi/2}e^{j2\pi(f_i\tau_{p,q}(t)-(f_i+\Delta f)\tau_{p',q'}(t+\Delta t))}\\&f(\theta_{p,c}^{\text{S}_\text{T}})f(\phi_{p,c}^{\text{S}_\text{T}})f(\theta_{q,c}^{\text{S}_\text{R}})f(\phi_{q,c}^{\text{S}_\text{R}})d\theta_{p,c}^{\text{S}_\text{T}}d\phi_{p,c}^{\text{S}_\text{T}}d\theta_{q,c}^{\text{S}_\text{R}}d\phi_{q,c}^{\text{S}_\text{R}}.
\end{split}
\end{equation}

--In the MB case,
\begin{equation}
\begin{split}
&R_{h,k}^\text{MB}(p,q,p',q',t,\Delta t,f_i, \Delta f)=\frac{P_k^\text{MB}}{K+1}\\&\int_{-\pi}^{\pi}\int_{-\pi/2}^{\pi/2}\int_{-\pi}^{\pi}\int_{-\pi/2}^{\pi/2}e^{j2\pi(f_i\tau_{p,q}(t)-(f_i+\Delta f)\tau_{p',q'}(t+\Delta t))}\\&f(\theta_{p,k}^{\text{M}_\text{T}})f(\phi_{p,k}^{\text{M}_\text{T}})f(\theta_{q,k}^{\text{M}_\text{R}})f(\phi_{q,k}^{\text{M}_\text{R}})d\theta_{p,k}^{\text{M}_\text{T}}d\phi_{p,k}^{\text{M}_\text{T}}d\theta_{q,k}^{\text{M}_\text{R}}d\phi_{q,k}^{\text{M}_\text{R}}.
\end{split}
\end{equation}
	
\subsection{Time Variant ACF}
The time-variant ACF reflects the correlation between two channels at different instants. It can be calculated by setting $p'=p$, $q'=q$ and $\Delta f = 0$ in (\ref{STFCF}) and can be expressed as 	
\begin{equation}
\begin{split}
&r_h(p,q,t,\Delta t,f_i)=R_h(p,q,t,\Delta t,f_i)=r_h^\text{LoS}(p,q,t,\Delta t,f_i)\\+&\sum_{c=1}^{C_\text{SB}}r_{h,c}^\text{SB}(p,q,t,\Delta t,f_i)+\sum_{k=1}^{C_\text{MB}}r_{h,k}^\text{MB}(p,q,t,\Delta t,f_i).
\end{split}
\end{equation}	

--In the LoS case,
\begin{equation}
\begin{split}
&r_h^\text{LoS}(p,q,t,\Delta t,f_i)=\frac{K}{K+1}e^{j2\pi f_i(\tau_{p,q}(t)-\tau_{p,q}(t+\Delta t))}.
\end{split}
\end{equation}

--In the SB case,
\begin{equation}
\begin{split}
&r_{h,c}^\text{SB}(p,q,t,\Delta t,f_i)=\frac{P_c^\text{SB}}{K+1}\\&\int_{-\pi}^{\pi}\int_{-\pi/2}^{\pi/2}\int_{-\pi}^{\pi}\int_{-\pi/2}^{\pi/2}e^{j2\pi f_i(\tau_{p,q}(t)-\tau_{p,q}(t+\Delta t))}\\&f(\theta_{p,c}^{\text{S}_\text{T}})f(\phi_{p,c}^{\text{S}_\text{T}})f(\theta_{q,c}^{\text{S}_\text{R}})f(\phi_{q,c}^{\text{S}_\text{R}})d\theta_{p,c}^{\text{S}_\text{T}}d\phi_{p,c}^{\text{S}_\text{T}}d\theta_{q,c}^{\text{S}_\text{R}}d\phi_{q,c}^{\text{S}_\text{R}}.
\end{split}
\end{equation}

--In the MB case,
\begin{equation}
\begin{split}
&r_{h,k}^\text{MB}(p,q,t,\Delta t,f_i)=\frac{P_k^\text{MB}}{K+1}\\&\int_{-\pi}^{\pi}\int_{-\pi/2}^{\pi/2}\int_{-\pi}^{\pi}\int_{-\pi/2}^{\pi/2}e^{j2\pi f_i(\tau_{p,q}(t)-\tau_{p,q}(t+\Delta t))}\\&f(\theta_{p,k}^{\text{M}_\text{T}})f(\phi_{p,k}^{\text{M}_\text{T}})f(\theta_{q,k}^{\text{M}_\text{R}})f(\phi_{q,k}^{\text{M}_\text{R}})d\theta_{p,k}^{\text{M}_\text{T}}d\phi_{p,k}^{\text{M}_\text{T}}d\theta_{q,k}^{\text{M}_\text{R}}d\phi_{q,k}^{\text{M}_\text{R}}.
\end{split}
\end{equation}

\subsection{Spatial CCF}
From (\ref{STFCF}), the spatial CCF can be derived by imposing $\Delta t = 0$ and $\Delta f = 0$ and can be written as 	
\begin{equation}
\begin{split}
&\rho_h(p,q,p',q',t,f_i)=R_h(p,q,p',q',t,f_i)\\=&\rho_h^\text{LoS}(p,q,p',q',t,f_i)+\sum_{c=1}^{C_\text{SB}}\rho_{h,c}^\text{SB}(p,q,p',q',t,f_i)\\&+\sum_{k=1}^{C_\text{MB}}\rho_{h,k}^\text{MB}(p,q,p',q',t,f_i).
\end{split}
\end{equation}	

--In the LoS case,
\begin{equation}
\begin{split}
&\rho_h^\text{LoS}(p,q,p',q',t,f_i)=\frac{K}{K+1}e^{j2\pi f_i(\tau_{p,q}(t)-\tau_{p',q'}(t))}.
\end{split}
\end{equation}

--In the SB case,
\begin{equation}
\begin{split}
&\rho_{h,c}^\text{SB}(p,q,p',q',t,f_i)=\frac{P_c^\text{SB}}{K+1}\\&\int_{-\pi}^{\pi}\int_{-\pi/2}^{\pi/2}\int_{-\pi}^{\pi}\int_{-\pi/2}^{\pi/2}e^{j2\pi f_i(\tau_{p,q}(t)-\tau_{p',q'}(t))}\\&f(\theta_{p,c}^{\text{S}_\text{T}})f(\phi_{p,c}^{\text{S}_\text{T}})f(\theta_{q,c}^{\text{S}_\text{R}})f(\phi_{q,c}^{\text{S}_\text{R}})d\theta_{p,c}^{\text{S}_\text{T}}d\phi_{p,c}^{\text{S}_\text{T}}d\theta_{q,c}^{\text{S}_\text{R}}d\phi_{q,c}^{\text{S}_\text{R}}.
\end{split}
\end{equation}

--In the MB case,
\begin{equation}
\begin{split}
&\rho_{h,k}^\text{MB}(p,q,p',q',t,f_i)=\frac{P_k^\text{MB}}{K+1}\\&\int_{-\pi}^{\pi}\int_{-\pi/2}^{\pi/2}\int_{-\pi}^{\pi}\int_{-\pi/2}^{\pi/2}e^{j2\pi f_i(\tau_{p,q}(t)-\tau_{p',q'}(t))}\\&f(\theta_{p,k}^{\text{M}_\text{T}})f(\phi_{p,k}^{\text{M}_\text{T}})f(\theta_{q,k}^{\text{M}_\text{R}})f(\phi_{q,k}^{\text{M}_\text{R}})d\theta_{p,k}^{\text{M}_\text{T}}d\phi_{p,k}^{\text{M}_\text{T}}d\theta_{q,k}^{\text{M}_\text{R}}d\phi_{q,k}^{\text{M}_\text{R}}.
\end{split}
\end{equation}

\subsection{FCF}
The FCF is another important channel statistical property reflecting the  correlation of the channels at different frequencies. It can be calculated by
\begin{equation}
\begin{split}
&T_h(p,q,t,f,\Delta f)\\=&E\left[H_{p,q}(t,f)\cdot H^*_{p,q}(t,f+\Delta f) \right]\\
=&E\left[ \int_{-\infty}^{\infty}{h}_{p,q}(t,\tau,f)e^{-j2\pi f\tau}d\tau \right.\\ &\left.\int_{-\infty}^{\infty}{h}_{p,q}(t,\tau,f+\Delta f)e^{-j2\pi f\tau}d\tau \right].
\end{split}
\end{equation}


\section{The Simulation Model and Statistical Properties}
\subsection{Description of the Simulation Model}
In the aforementioned theoretical channel model, each cluster is composed of an infinite number of rays, which can not be implemented in hardware or software. According to the theoretical model, the simulation model comprises finite rays in each cluster expressed as
\begin{equation}\label{SM_CIR}
\begin{split}
\hspace{3mm}
\tilde{h}_{p,q}(t,\tau,f_{i})=&\tilde{h}_{p,q}^{\text{LoS}}(t,\tau,f_{i})+
\sum_{c=1}^{C_\text{SB}}\tilde{h}_{p,q,c}^{\text{SB}}(t,\tau,f_{i})\\&+\sum_{k=1}^{C_\text{MB}}\tilde{h}_{p,q,k}^{\text{MB}}(t,\tau,f_{i})
\end{split}
\end{equation} 
where the LoS part is expressed as 
\begin{equation}\label{SM_LoS}
\begin{split}
\tilde{h}_{p,q}^{\text{LoS}}(t,\tau,f_{i})=&\sqrt{\frac{K}{K+1}}\times e^{-j(2\pi f_{i}\tau_{p,q}^{\text{LoS}}(t))}\delta(\tau-\tau_{p,q}^{\text{LoS}}(t)). 
\end{split}
\end{equation}

The SB and MB components of the simulation model are 
\begin{equation}\label{SM_SB}
\begin{split}
\tilde{h}_{p,q,c}^{\text{SB}}(t,\tau,f_{i})=&\sqrt{\frac{P^\text{SB}_c}{K+1}}\ \sum_{l=1}^{L_c} \sqrt{\frac{1}{L_c}}\\&e^{-j(2\pi f_{i}\tau_{p,q,c,l}^{\text{SB}}(t)-\Theta^\text{SB}_{c,l})}\delta(\tau-\tau_{p,q,c,l}^{\text{SB}}(t))
\end{split}
\end{equation}
\begin{equation}\label{SM_MB}
\begin{split}
\tilde{h}_{p,q,k}^{\text{MB}}(t,\tau,f_{i})&=\sqrt{\frac{P^\text{MB}_k}{K+1}}\sum_{m=1}^{M_k} \sqrt{\frac{1}{M_k}}\\&e^{-j(2\pi f_{i}\tau_{p,q,c,m,n}^{\text{MB}}(t)-\Theta^\text{MB}_{c,m})}\delta(\tau-\tau_{p,q,c,m}^{\text{MB}}(t)).
\end{split}
\end{equation}

In this simulation model,  relative angles are discrete for simulation. In each cluster, maximum relative angle corresponds to maximum relative delay so that the  discrete relative delay can be generated by relative angles.  Hence, an appropriate parameter calculation method is necessary to approximate the properties of the theoretical channel model. In this simulation model, we apply the MEA \cite{Mobile} method to generate the discrete angles. {In this simulation model, the MEA method is used to calculate limited discrete angles according to principle that the integral of the probability between two adjacent angles are equal. } 

Here, we use $\theta_{p,c,l}^{\text{S}_\text{T}}$  as an example to generate $\sqrt{L_c}$ discrete values with the MEA method. The whole probability are divided into $\sqrt{L_c}$ parts with the same area $1/\sqrt{L_c}$. According to the equation $\int_{l-1}^{l}dF(x)=1/\sqrt{L_c}$ where $F(x)$ is the cumulative distribution function (CDF) of $\theta_{p,c,l}^{\text{S}_\text{T}}$, then $\sqrt{L_c}$ discrete angle can be obtained. Similarly, the  $\phi_{p,c,l}^{\text{S}_\text{T}}$
can also generate  $\sqrt{L_c}$ discrete values. So there are $L_c$ discrete values in this cluster. Meanwhile, the discrete $\theta_{q,c,l}^{\text{S}_\text{R}}$ and $\phi_{q,c,l}^{\text{S}_\text{R}}$  can also be obtained.

\subsection{Statistical Properties of the Simulation Model}
In this sub-section, the statistical properties of the  non-stationary THz simulated model are derived. 



\subsubsection{STFCF}
The STFCF of the simulation model
$\tilde{R}_{h}(p,q,p',q',t,\Delta t,f,\Delta f)$ can be reppresented as
\begin{equation}
\label{STF_correlation_function}
\begin{split}
&\tilde{R}_{h}(p,q,p',q',t,\Delta t,f,\Delta f)\\
= &E\left[\tilde{h}_{p,q}(t,f_i)\cdot \tilde{h}_{p',q'}^*(t+\Delta t,f+\Delta f) \right] \\
=&\tilde{R}_h^\text{LoS}(p,q,p',q',t,\Delta t,f_i, \Delta f)\\&+\sum_{c=1}^{C_\text{SB}}\tilde{R}_{n,c}^\text{SB}(p,q,p',q',t,\Delta t,f_i, \Delta f) \\&+\sum_{k=1}^{C_\text{MB}}\tilde{R}_{h,k}^\text{MB}(p,q,p',q',t,\Delta t,f_i, \Delta f).
\end{split}
\end{equation}
The correlation function of the channel also consists of the LoS, SBCs, and MBCs. 

--In the LoS case,
\begin{equation}
\begin{split}
&\tilde{R}_h^\text{LoS}(p,q,p',q',t,\Delta t,f_i, \Delta f)=\\&\frac{K}{K+1}e^{j2\pi(f_i\tau_{p,q}(t)-(f_i+\Delta f)\tau_{p',q'}(t+\Delta t))}.
\end{split}
\end{equation}

--In the SB case
\begin{equation}
\begin{split}
&\tilde{R}_{h,c}^\text{SB}(p,q,p',q',t,\Delta t,f_i, \Delta f)=\frac{P_c^\text{SB}}{K+1}\\&\sum_{l=1}^{L_c}e^{j2\pi(f_i\tau_{p,q}(t)-(f_i+\Delta f)\tau_{p',q'}(t+\Delta t))}.
\end{split}
\end{equation}

--In the MB case,
\begin{equation}
\begin{split}
&\tilde{R}_{h,k}^\text{MB}(p,q,p',q',t,\Delta t,f_i, \Delta f)=\frac{P_k^\text{MB}}{K+1}\\&\sum_{m=1}^{M_k}e^{j2\pi(f_i\tau_{p,q}(t)-(f_i+\Delta f)\tau_{p',q'}(t+\Delta t))}.
\end{split}
\end{equation}
By setting partial parameters of ($\Delta p,\Delta q,\Delta t$, $\Delta f$) as 0, the STFCF can easily be simplified to time-variant ACF, spatial CCF, and FCF.

\subsubsection{The Delay PSD}
The delay PSD of the simulation model $\Upsilon_{p,q}(t,\tau,f_{i})$ is calculated as 
\begin{equation}\label{PDP}
\begin{split}
&\Upsilon_{p,q}(t,\tau,f_{i})=
\left| \tilde{h}_{p,q}(t,\tau,f_{i})\right|^2\\&
=\left| \tilde{h}^\text{LoS}_{p,q}(t,\tau,f_{i})\right|^2+\sum_{c=1}^{C_\text{SB}}\left|\tilde{h}_{p,q,c}^{\text{SB}}(t,\tau,f_{i})\right|^2\\&+\sum_{k=1}^{C_\text{MB}}\left|\tilde{h}_{p,q,k}^{\text{MB}}(t,\tau,f_{i})\right|^2.
\end{split}
\end{equation}

It should be noted that the time-space-frequency variant properties of delay PSD is caused by the time-space-frequency variant powers and delays. 

\subsubsection{The Stationary Intervals}

The stationary interval is the time, distance, and frequency bandwidth during which the propagation channel can be considered as unchanged in time, space, and frequency domains, respectively. The stationary interval is defined as the maximum length within which the ACF of the time-space-frequency variant delay PSD exceeds the threshold \cite{RN516}. 
Here we calculate the stationary interval in time and frequency domains expressed as $I(\Delta t, \Delta f)=\text{inf}\left\lbrace \Delta t, \Delta f\mid_{R_\Upsilon(t,f,\Delta t, \Delta f)<c_{th}}\right\rbrace$ where $\text{inf} \left\lbrace \cdot \right\rbrace $ calculates the infimum of a function. The    normalized ACF of the delay PSD $R_\Lambda(t,f,\Delta t, \Delta f)$ is defined by
\begin{equation}\label{stationary_bandwidth}
\begin{split}
&R_\Lambda(t,f_i,\Delta t, \Delta f)=\\&\frac{\int \Upsilon_{p,q}(t,\tau,f_{i})\Upsilon_{p,q}(t_i+\Delta t,\tau,f_{i}+\Delta f)d\tau}{max\left\lbrace \int \Upsilon_{p,q}^2(t,\tau,f_{i}) d\tau, \int \Upsilon_{p,q}^2(t_i+\Delta t,\tau,f_{i}+\Delta f) d\tau\right\rbrace}.
\end{split}
\end{equation}



\section{Results and Analysis}
In this section,  the statistical properties of the THz channel models are simulated and discussed. The frequency band is chosen from 300~GHz to 350~GHz. The numbers of antenna elements at the Tx and Rx are 256 at both Tx and Rx. Both $\delta_\text{T}$ and $\delta_\text{R}$ equal to half of wavelength at 325 GHz. The moving speed of Rx is $\text{v}^\text{R}$ = 0.1 m/s with the direction angle $\zeta^\text{R}$ = 0 and $\xi^\text{R}$~=~$\frac{\pi}{3}$ while the Tx is fixed. The bandwidth of sub-band is 0.1 GHz. It should be noted that it is less than the real stationary bandwidth to make sure that each sub-band in the simulation is frequency stationary. The number of rays in each cluster is  $L_c =  M_c = 400$.  

%

\subsection{ACF}
\begin{figure}[t]
	\centerline{\includegraphics[width=0.47\textwidth]{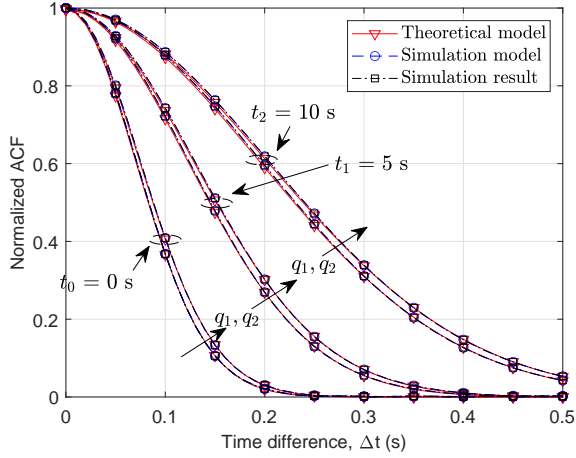}}
	\caption{The comparison of time-variant ACFs of theoretical model, simulation model, and simulation results  for cluster $C_1^\text{S}$ at $t_0$ = 0 s, $t_1$ = 5 s and $t_2$ = 10 s with $q_1$ = 1 and $q_2$ = 200 ($D = 3$ m, $p$ = 1, $f_0$ = 325 GHz, $\text{v}^\text{R}$ = 0.1 m/s, $\zeta^\text{R}$ = 0, $\xi^\text{R}$~=~$\frac{\pi}{3}$, $\mu_{c,\theta}^{\text{S}_\text{R}}$ = $1.4^\circ$, $\mu_{c,\phi}^{\text{S}_\text{R}}$ = $2.8^\circ$, $r^{\text{S}_\text{T}}_{c}$ = 0.4).}
	\label{fig_ACF}
\end{figure}

The time ACF of the model can be calculated by setting $\Delta p,\Delta q,\Delta f$ as 0.
The comparison of theoretical model, simulation model, and simulation result at  $t_0$ = 0 s, $t_1$ = 5 s, and $t_2$ = 10~s of $\text{Cluster}_1$ at different antennas is shown in Fig.~\ref{fig_ACF}. From the results, we can observe that the simulation models fit well with the theoretical model and the simulation results. We can also observe different time ACFs at different time instants, showing the non-stationarity in time domain. The difference between different {receive antennas} can also be observed clearly showing the non-stationarity of ultra-massive MIMO. 

\subsection{Spatial CCF}
\begin{figure}[tb]
	\centerline{\includegraphics[width=0.45\textwidth]{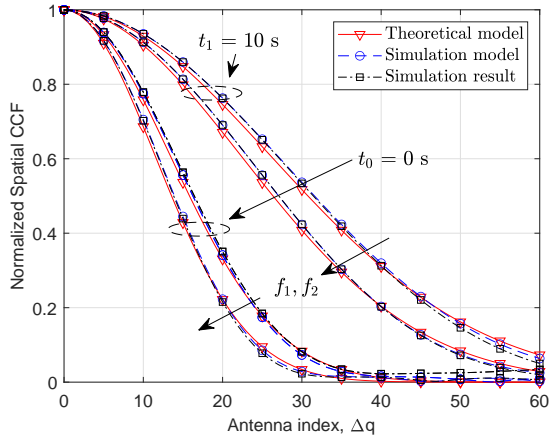}}
	\caption{The comparison of spatial CCFs of theoretical model, simulation model, and simulation results  for cluster $C_1^\text{S}$ ($D = 3$ m, $p$ = 1, $q$ = 1, $t_0$ = 0 s,  $t_1$ = 10 s, $f_1$ = 300 GHz, $f_2$ = 350 GHz, $\text{v}^\text{R}$ = 0.1 m/s, $\zeta^\text{R}$ = 0, $\xi^\text{R}$~=~$\frac{\pi}{3}$,  $\alpha^\text{R} = \frac{\pi}{3}$, $\beta^\text{R} = \frac{\pi}{4}$,  $\mu_{c,\theta}^{\text{S}_\text{R}}$ = $1.4^\circ$, $\mu_{c,\phi}^{\text{S}_\text{R}}$ = $2.8^\circ$, $r^{\text{S}_\text{T}}_{c}$ = 0.4).}
	\label{fig_CCF}
\end{figure}

The spatial CCFs of the theoretical model, simulation model, and simulation result for $\text{Cluster}_1$ at different time and different frequencies are shown in Fig. \ref{fig_CCF}. We can see that the theoretical model, simulation model, and simulation result match very well. {We can also observe significant differences at different time instants. Due to the traveling of the Rx, the channel shows non-stationarity in the time domain. In addition, we present the results in different frequencies showing unnegligible frequency non-stationarity. For higher frequencies, the spatial CCFs tend to be smaller.}
\subsection{FCF}
\begin{figure}[t]
	\centerline{\includegraphics[width=0.47\textwidth]{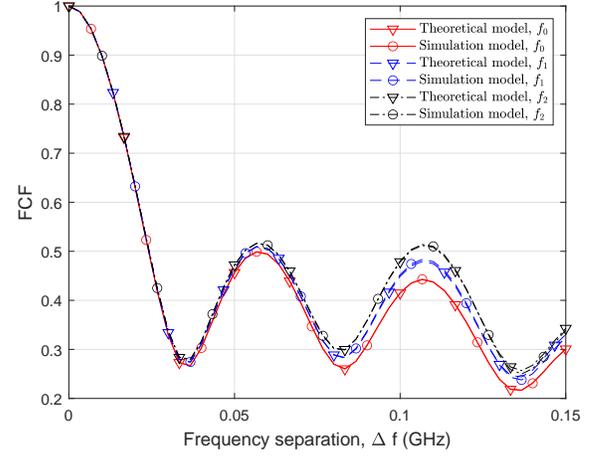}}
	\caption{The comparison of FCFs of theoretical model and simulation model at $f_0$ = 300 GHz, $f_1$ = 325 GHz and $f_2$ = 350 GHz ($D = 3$ m, $p$ = 1, $q$ = 1, $t_0$ = 0~s, $\text{v}^\text{R}$ = 0 m/s, $\mu_{c,\theta}^{\text{S}_\text{T}}$ = $1.2^\circ$, $\mu_{c,\phi}^{\text{S}_\text{T}}$ = $1.7^\circ$, $\mu_{c,\theta}^{\text{S}_\text{R}}$ = $1.4^\circ$, $\mu_{c,\phi}^{\text{S}_\text{R}}$ = $2.8^\circ$, $\mu^\text{ToA}_c$ = 0.3 ns, $\rho_{\phi}^{\text{S}_\text{T}} = \rho_{\varphi}^{\text{S}_\text{T}} = \rho_{\phi}^{\text{S}_\text{R}} = \rho_{\varphi}^{\text{S}_\text{R}}$ = 1.2).}
	\label{fig_FCF}
\end{figure}

The absolute values of FCFs for NLoS paths at different  $f_0$ = 300 GHz, $f_1$ = 325 GHz, and $f_2 = 350$ GHz for THz massive MIMO channel model are shown in Fig.~\ref{fig_FCF}. We can notice that differences among FCFs at different frequencies are small but still observable which means that  the non-stationarity in the frequency domain needs to be considered. 
{When frequency difference increases, the gaps between different frequencies will also increase because the change rate of frequency correlation is related to the center frequency.}

\subsection{Stationary Bandwidth}
\begin{figure}[t]
	\centerline{\includegraphics[width=0.47\textwidth]{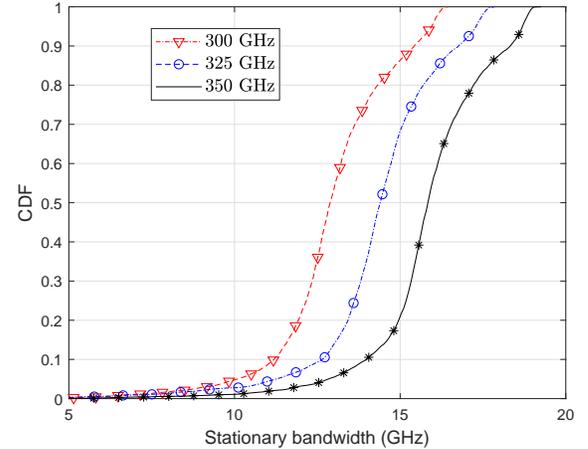}}
	\caption{The CDFs of stationary interval of the simulation model in frequency domain at $f_0$ = 300 GHz, $f_1$ =325 GHz and $f_2$ = 350 GHz ($D = 3$ m, $p$ = 1, $q$ = 1, $t_0$ = 0~s, $\text{v}^\text{R}$ = 0.1 m/s, $\mu_{c,\theta}^{\text{S}_\text{T}}$ = $1.2^\circ$, $\mu_{c,\phi}^{\text{S}_\text{T}}$ = $1.7^\circ$, $\mu_{c,\theta}^{\text{S}_\text{R}}$ = $1.4^\circ$, $\mu_{c,\phi}^{\text{S}_\text{R}}$ = $2.8^\circ$, $\mu^\text{ToA}_c$ = 0.3 ns, $\rho_{\phi}^{\text{S}_\text{T}} = \rho_{\varphi}^{\text{S}_\text{T}} = \rho_{\phi}^{\text{S}_\text{R}} = \rho_{\varphi}^{\text{S}_\text{R}}$ = 1.2, $c_{th} = 0.9$).}
	\label{Stationinterval}
\end{figure}
The stationary bandwidth is simulated with parameters of a typical indoor scenario according to\cite{RN194,RN504}. The initial distance between the first elements of Tx and Rx is 3 m, $\mu_{\Delta \tau_{c,\text{SB}}}$ and $\mu_{\Delta \tau_{k,\text{MB}}}$ are set as 2.73 ns and 2.33 ns. 
The initial intra-cluster parameters $\mu_{c,\theta}^{\text{S}_\text{T}}$, $\mu_{c,\phi}^{\text{S}_\text{T}}$,  $\mu_{c,\theta}^{\text{S}_\text{R}}$, and $\mu_{c,\phi}^{\text{S}_\text{R}}$ are set as $1.2^\circ$, $1.7^\circ$, $1.4^\circ$, and $2.8^\circ$, respectively. The mean value of relative time of arrival $\mu^{ToA}_c$ is set as 0.3 ns. The CDFs of the stationary bandwidth of the simulation channel model at different frequencies are shown in Fig.~\ref{Stationinterval}. 
The threshold is chosen as 0.9. 
The median of the stationary bandwidth at 300 GHz is approximately 12.5~GHz. Higher frequency band has larger frequency stationary bandwidth. If the bandwidth in THz communication is large than the stationary bandwidth, non-stationarity in frequency domain is unneglectable.

\subsection{Cluster Level Angle Spread}
\begin{figure}[t]
	\centerline{\includegraphics[width=0.45\textwidth]{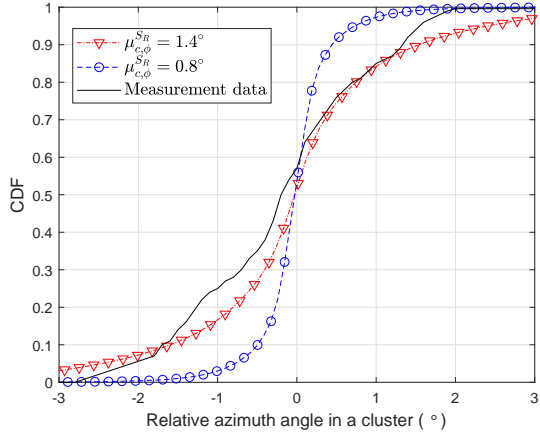}}
	\caption{The CDFs of relative angle in a cluster with different $\mu_{c,\phi}^{\text{S}_\text{R}}$ and the measurement in \cite{RN197} ($D = 2.7$ m, $p$ = 1, $q$ = 1, $t_0$ = 0 s, $f_0$ = 300 GHz, $\text{v}^\text{R}$ = 0 m/s).}
	\label{relative_delay}
\end{figure}

\begin{figure}[t]
	\centerline{\includegraphics[width=0.47\textwidth]{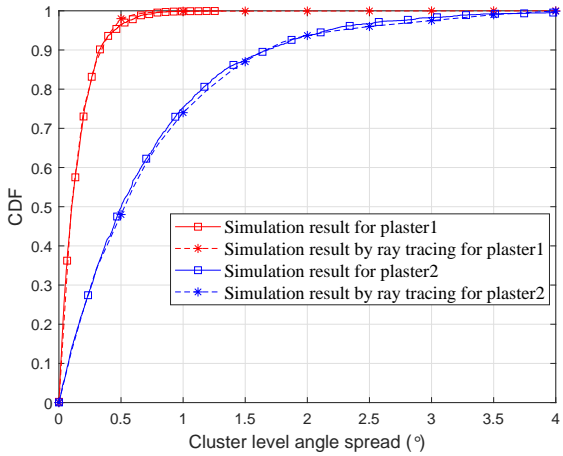}}
	\caption{The comparison of cluster level angular spread with different materials ($p$ = 1, $q$ = 1, $t_0$ = 0 s, $f_0$ = 300 GHz, $\text{v}^\text{R}$ = 0 m/s, $r^{\text{S}_\text{T}}_{c}$ = 0.4, $\mu_{c,\phi}^{\text{S}_\text{R}}$ = $0.15^\circ$ for  plaster1, $\mu_{c,\phi}^{\text{S}_\text{R}}$ = $0.75^\circ$ for  plaster2).}
	\label{Cluster_level_Anglespread}
\end{figure}
Cluster level angle spread reflect the degree of diffusely scattering in a cluster. 
In this simulation, the parameter $\mu_{c,\phi}^{\text{S}_\text{R}}$ 
are estimated by using an optimization algorithm in order to fit the statistical properties of the channel model to those of the data from  measurement \cite{RN197} or ray tracing \cite{RN195}. {In the  curve fitting (optimization) process, random initial values of those parameters were initialized.  Then, the average mean square error of the simulation and measurement results is minimized by optimizing the values of those parameters in an iterative procedure. After the sufficient numbers of iterations, the optimal values of those parameters with the minimum mean square error can be found, which allows the desirable good fitting between the statistical property of the simulation model and measurement data.} 
Firstly, the CDFs of relative azimuth angles of arrival with different $\mu_{c,\phi}^{\text{S}_\text{R}}$ are simulated and compared with the measurement data\cite{RN197} in Fig. \ref{relative_delay}. {When the parameter $\mu_{c,\phi}^{\text{S}_\text{R}}$ increases, the CDF of relative angles will be more gentle. We can observe the good agreement when $\mu_{c,\phi}^{\text{S}_\text{R}}=1.4^\circ$ indicating that the modeling methods can be applied to simulate realistic environments.}  Then we simulate the scattering in two materials with different roughness. The cluster level angle spread is greatly affected by the roughness of the material. The height standard deviation of plaster1 is $\sigma_{\text{plaster1}}=0.5$ mm and $\sigma_{\text{plaster2}}=1.5$~mm in \cite{RN195}.  Fig.~\ref{Cluster_level_Anglespread} demonstrates the comparison of cluster level angle spread with different materials in this  model and simulation by ray tracing in \cite{RN195}. {The results indicate that the proposed model can correctly approximate the results from ray tracing for different materials.}


\subsection{RMS Delay Spread}

\begin{figure}[t]
	\centerline{\includegraphics[width=0.45\textwidth]{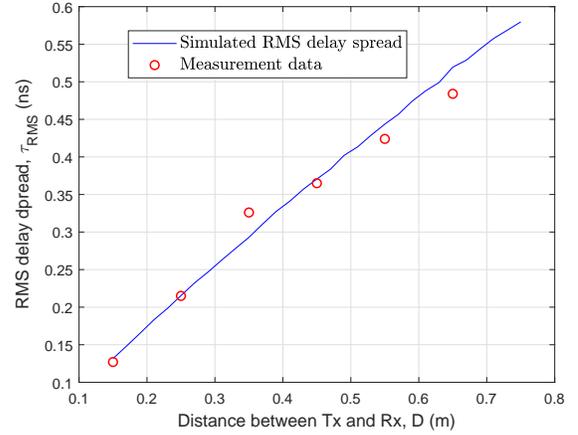}}
	\caption{The comparison of simulated RMS delay spread and measurement in \cite{RN448} ($p$ = 1, $q$ = 1, $t_0$ = 0 s, $f_0$ = 300 GHz, $\text{v}^\text{R}$ = 0~m/s, $\mu_{\Delta \tau_{\text{SB}}}$ = 0.08,  $\mu_{\Delta \tau_{\text{MB}}}$ = 0.07 for $D$ = 0.15 m).}
	\label{rms}
\end{figure}
The RMS delay spread is simulated and fitted with the measurement data\cite{RN448} by minimum mean square error criterion in Fig. \ref{rms} for short range communication where the transmission distance is less than 1 m. In the simulation, the parameters $\mu_{\Delta \tau_{c,\text{SB}}}$ and $\mu_{\Delta \tau_{k,\text{MB}}}$ are linearly increase with the LoS distance and the parameters at $D$ = 0.15 m are estimated. We can observe good agreement between the simulation results and the measurement data. It means that the proposed model is suitable for this scenario. 



\section{Conclusions}
Novel 3D space-time-frequency non-stationary ultra-massive MIMO THz channel models have been proposed for 6G wireless communication systems with long traveling paths. Considering the unique propagation characteristics of THz bands, i.e., frequency-dependent diffusely scattering, relative angles and delays of rays within one cluster have been assumed to be frequency variant parameters. The statistical properties such as ACF, spatial CCF, and FCF have been derived for the proposed theoretical model and corresponding simulation model based on the MEA. Numerical results have shown that the statistical properties of the simulation model, verified by simulation results, can match well with those of the theoretical model. The good agreements between simulated results and corresponding measurements in different scenarios have further demonstrated the accuracy and good flexibility of the proposed model.   



\begin{IEEEbiography}[{\includegraphics[width=1in,clip,keepaspectratio]{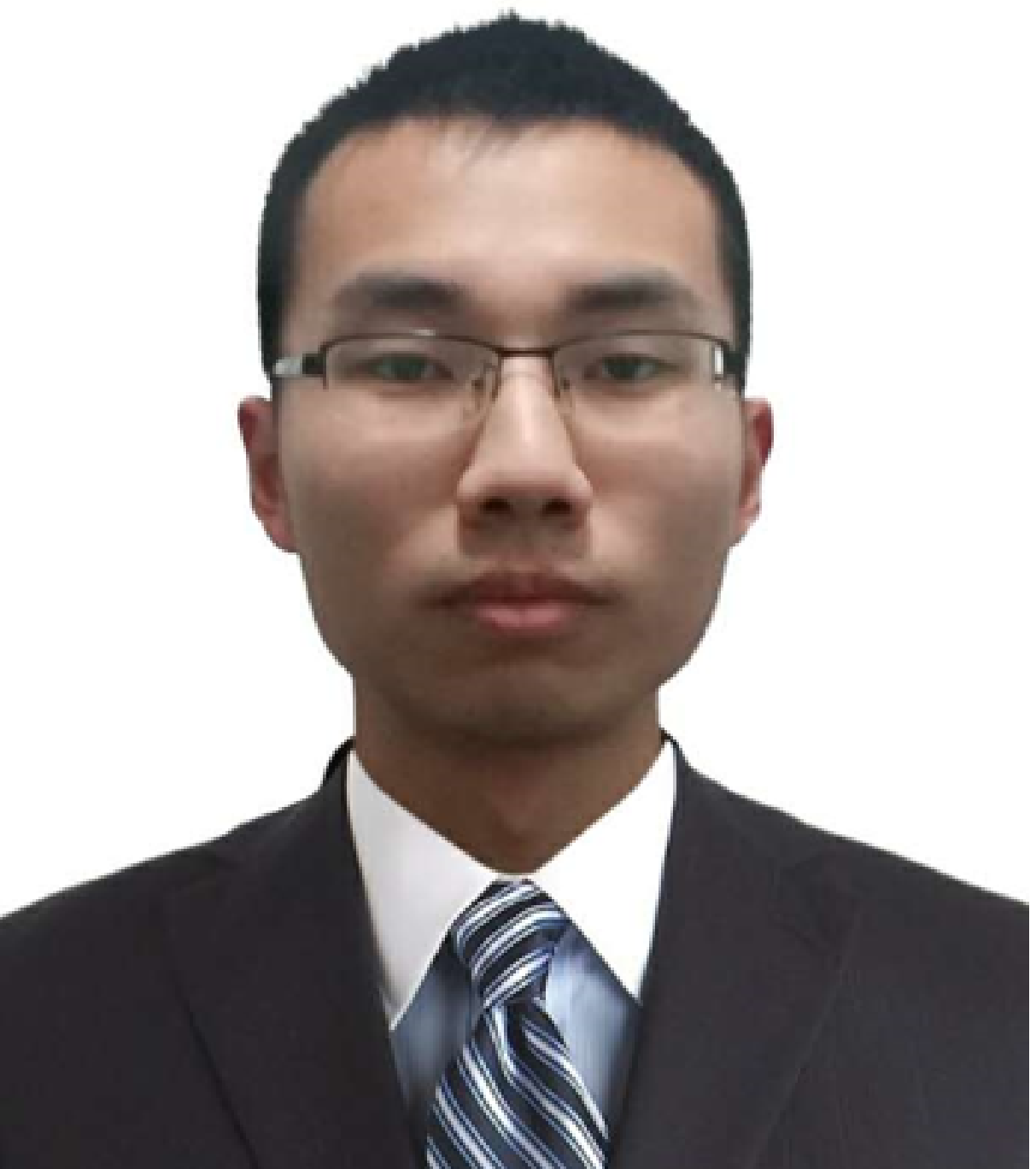}}]{Jun Wang}
	received the B.E. degree in Information Engineering from Southeast University, China, in 2016. He is currently pursuing the Ph.D. degree in the Nation Mobile
	Communications Research Laboratory, Southeast University, China. His research interests is THz wireless channel measurements and modeling.
\end{IEEEbiography}

\begin{IEEEbiography}[{\includegraphics[width=1in,height=1.25in,clip,keepaspectratio]{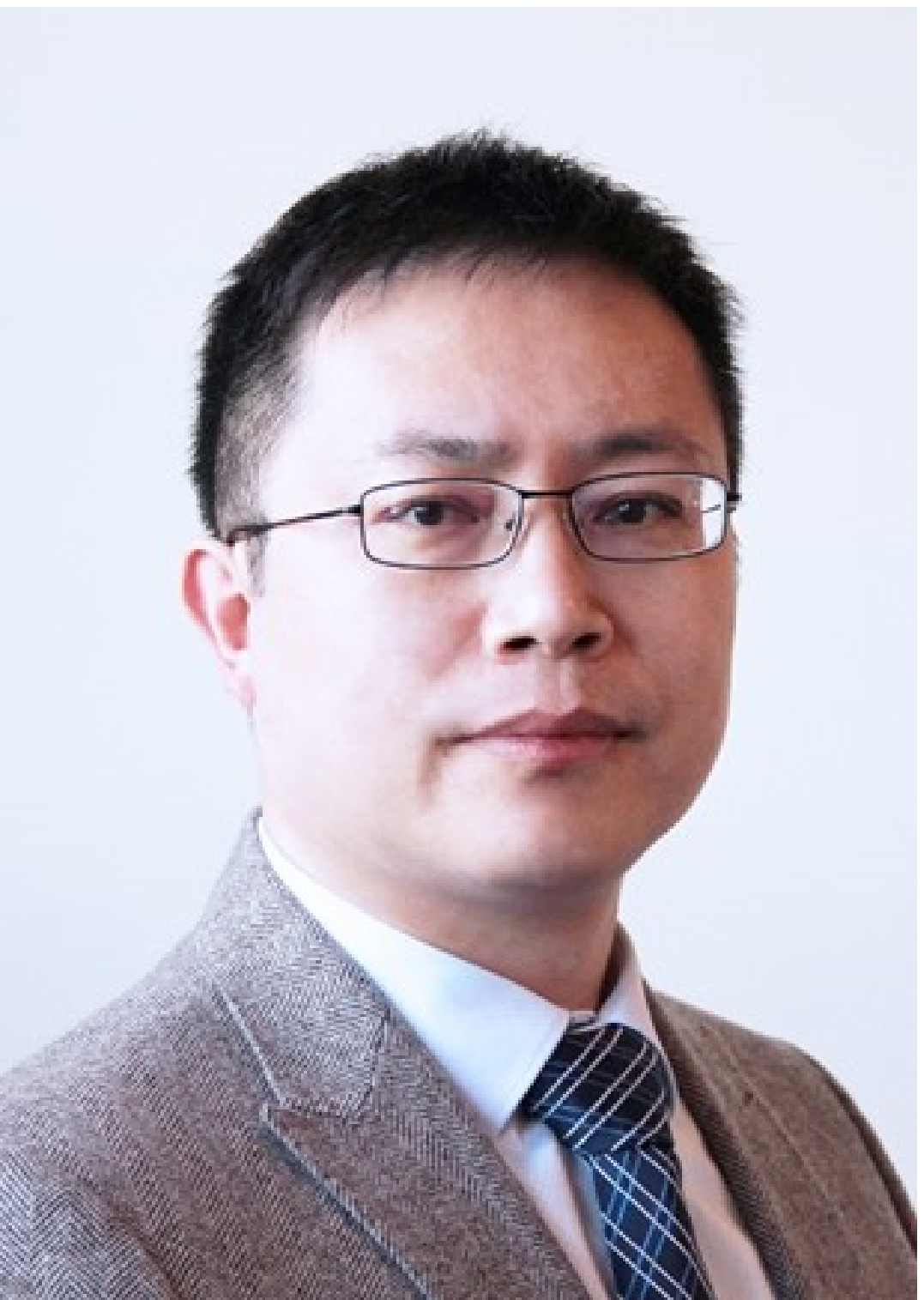}}]{Cheng-Xiang Wang}
	(S'01-M'05-SM'08-F'17) received the BSc and MEng degrees in
	Communication and Information Systems from	Shandong University, China, in 1997 and 2000,	respectively, and the PhD degree in Wireless Communications	from Aalborg University, Denmark, in 2004.
	
	He was a Research Assistant with the Hamburg University of Technology, Hamburg, Germany, from 2000 to 2001, a Visiting Researcher with
	Siemens AG Mobile Phones, Munich, Germany, in 2004, and a Research
	Fellow with the University of Agder, Grimstad, Norway, from 2001 to 2005. He has been with Heriot-Watt University, Edinburgh, U.K., since 2005, where he was promoted to a Professor in 2011. In 2018, he joined the National Mobile Communications Research Laboratory, Southeast University, China, as a Professor. He is also a part-time professor with the Purple Mountain Laboratories, Nanjing, China. He has authored four books, one book chapter, and more than 420 papers in refereed journals and conference proceedings, including 24 ESI Highly Cited Papers. He has also delivered 22  Invited Keynote Speeches/Talks and~7 Tutorials in international conferences. His current research interests include wireless channel measurements and modeling, B5G wireless communication networks, and	applying artificial intelligence to wireless networks.

	Prof. Wang is a Member of the Academia Europaea (The Academy of Europe), a fellow of the IET, an IEEE Communications Society
	Distinguished Lecturer in 2019 and 2020, and a Highly-Cited Researcher
	recognized by Clarivate Analytics in 2017-2020. He is currently an Executive Editorial Committee member for the IEEE TRANSACTIONS ON
	WIRELESS COMMUNICATIONS. He has served as an Editor for nine
	international journals, including the IEEE TRANSACTIONS ON WIRELESS
	COMMUNICATIONS from 2007 to 2009, the IEEE TRANSACTIONS
	ON VEHICULAR TECHNOLOGY from 2011 to 2017, and the
	IEEE TRANSACTIONS ON COMMUNICATIONS from 2015 to 2017.
	He was a Guest Editor for the IEEE JOURNAL ON SELECTED AREAS
	IN COMMUNICATIONS, Special Issue on Vehicular Communications and
	Networks (Lead Guest Editor), Special Issue on Spectrum and Energy
	Efficient Design of Wireless Communication Networks, and Special Issue
	on Airborne Communication Networks. He was also a Guest Editor for 	the IEEE TRANSACTIONS ON BIG DATA, Special Issue on Wireless 	Big Data, and is a Guest Editor for the IEEE TRANSACTIONS ON 	COGNITIVE COMMUNICATIONS AND NETWORKING, Special Issue on Intelligent Resource Management for 5G and Beyond. He received twelve Best Paper Awards from IEEE GLOBECOM 2010, IEEE ICCT 2011, ITST 2012, IEEE VTC 2013-Spring, IWCMC 2015, IWCMC 2016, IEEE/CIC ICCC 2016, WPMC 2016, WOCC 2019, IWCMC 2020 and WCSP 2020.
\end{IEEEbiography}

\begin{IEEEbiography}[{\includegraphics[width=1in,height=1.25in,clip,keepaspectratio]{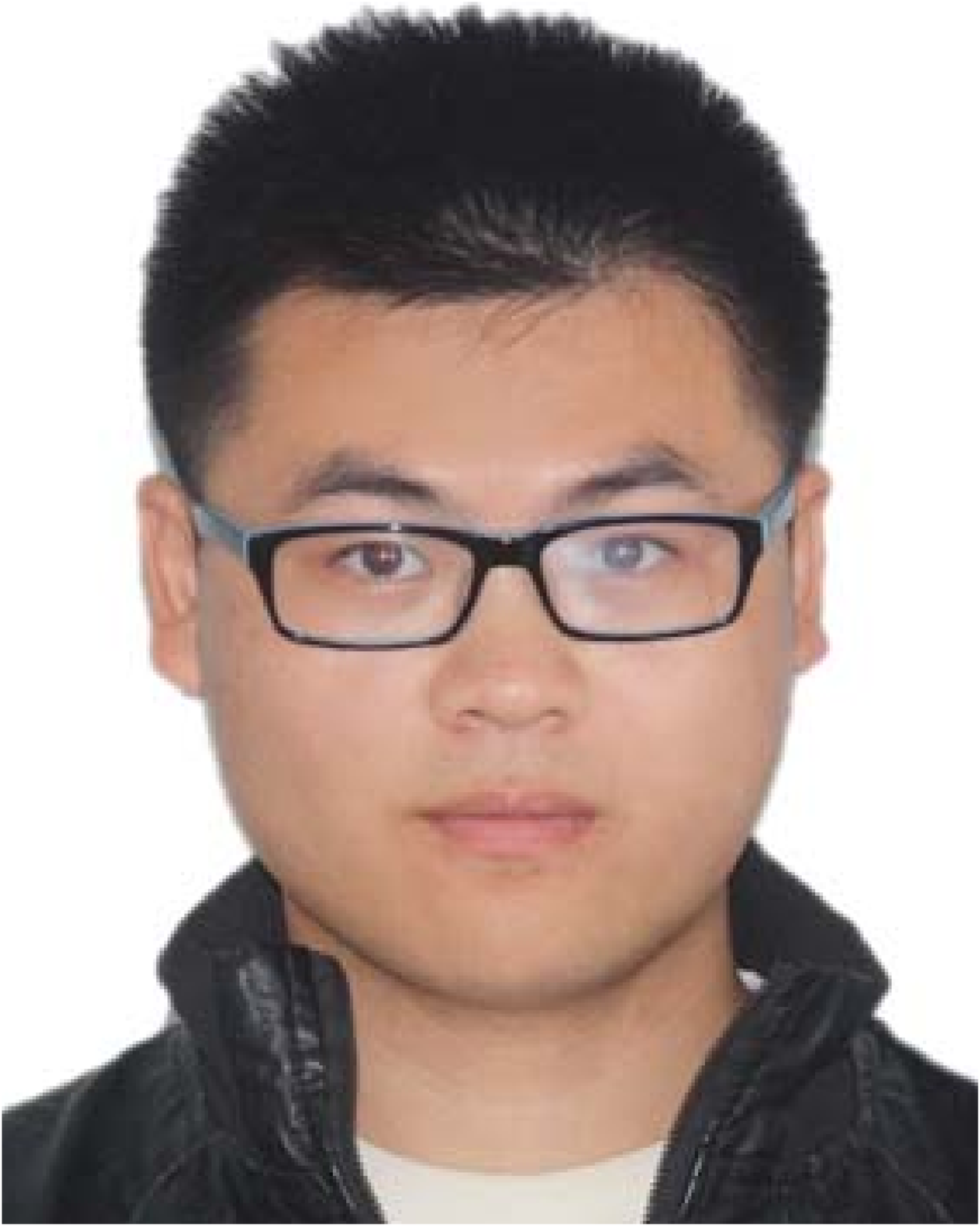}}]{Jie Huang}
(M'20) received the B.E. degree in Information Engineering from Xidian University, China, in 2013, and the Ph.D. degree in Communication and Information Systems from Shandong University, China, in 2018. From October 2018 to October 2020, he was a Postdoctoral Research Associate in the National Mobile Communications Research Laboratory, Southeast University, China, supported by the National Postdoctoral Program for Innovative Talents. From January 2019 to February 2020, he was a Postdoctoral Research Associate in Durham University, UK. He is currently an Associate Professor in the National Mobile Communications Research Laboratory, Southeast University, China and also a researcher in Purple Mountain Laboratories, China. His research interests include millimeter wave, THz, massive MIMO, intelligent reflecting surface channel measurements and modeling, wireless big data, and 6G wireless communications. He received Best Paper Awards from WPMC 2016 and WCSP 2020.

\end{IEEEbiography}

\begin{IEEEbiography}[{\includegraphics[width=1in,height=1.25in,clip,keepaspectratio]{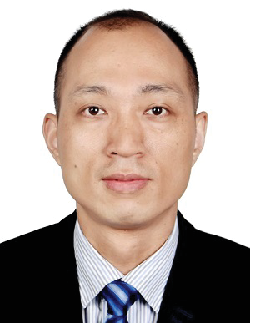}}]{Haiming Wang}
	(M'08) was born in 1975.
	He received the M.S. and Ph.D. degrees in electrical engineering from Southeast University, Nanjing,
	China, in 2002 and 2009, respectively.
	In 2002, he joined the State Key Laboratory of
	Millimeter Waves, Southeast University, where he is
	currently a Professor. In 2008, he was a Short-Term
	Visiting Scholar with the Blekinge Institute of Technology, Karlskrona, Sweden. He has authored and
	coauthored more than 50 technical publications in
	the IEEE TRANSACTIONS ON ANTENNAS AND
	PROPAGATION, the IEEE ANTENNAS AND WIRELESS PROPAGATION LETTERS, and other peer-reviewed academic journals. He has authored and
	coauthored more than 50 patents and 23 patents have been granted. His current
	research interests include millimeter-wave wireless mobile communications,
	millimeter-wave radar and imaging, radio propagation measurement and
	channel modeling, multiband and wideband antennas, and arrays.
	Dr. Wang was a recipient of the Science and Technology Progress Award of
	Jiangsu Province of China in 2009. In 2018, he was recognized by the IEEESA for his contributions to the development of IEEE 802.11aj. He served as
	the Vice Chair of the IEEE 802.11aj Task Group from 2012 to 2018.
\end{IEEEbiography}

\begin{IEEEbiography}[{\includegraphics[width=1in,height=1.25in,clip,keepaspectratio]{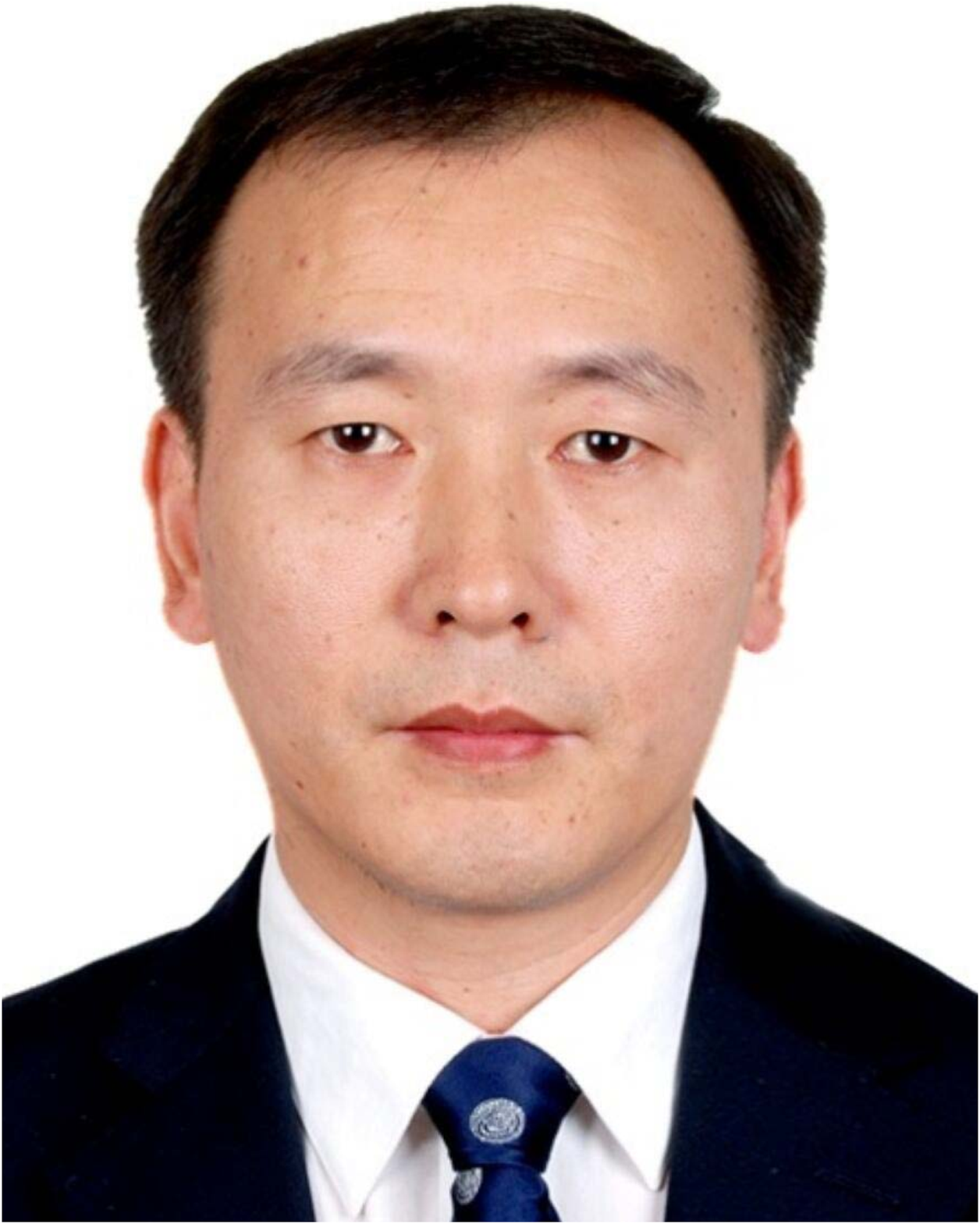}}]{Xiqi Gao}
	(S'92-AM'96-M'02-SM'07-F'15) received the Ph.D. degree in electrical engineering from Southeast University, Nanjing, China, in 1997. 
	
	Dr. Gao joined the Department of Radio Engineering, Southeast University, in April 1992. Since May 2001, he has been a professor of information systems and communications. From September 1999 to August 2000, he was a visiting scholar at Massachusetts Institute of Technology, Cambridge, MA, USA, and Boston University, Boston, MA. From August 2007 to July 2008, he visited the Darmstadt University of Technology, Darmstadt, Germany, as a Humboldt scholar. His current research interests include broadband multicarrier communications, MIMO wireless communications, channel estimation and turbo equalization, and multirate signal processing for wireless communications. From 2007 to 2012, he served as an Editor for the IEEE Transactions on Wireless Communications. From 2009 to 2013, he served as an Editor for the IEEE Transactions on Signal Processing. From 2015 to 2017, he served as an Editor for the IEEE Transactions on Communications. 
	
	Dr. Gao received the Science and Technology Awards of the State Education Ministry of China in 1998, 2006, and 2009, the National Technological Invention Award of China in 2011, and the 2011 IEEE Communications Society Stephen O. Rice Prize Paper Award in the field of communication theory.
\end{IEEEbiography}

\begin{IEEEbiography}[{\includegraphics[width=1in,height=1.25in,clip,keepaspectratio]{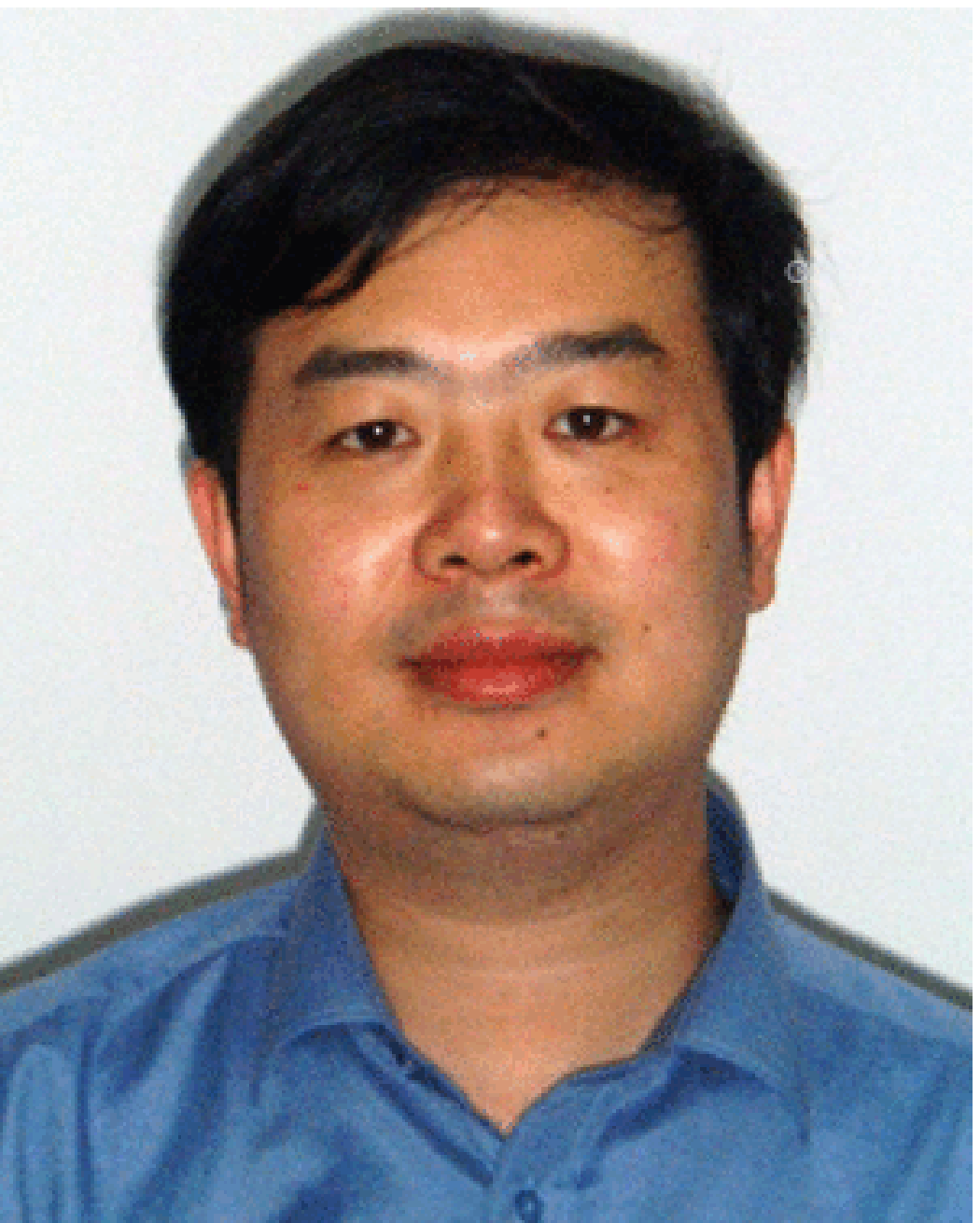}}]{Xiaohu You}
	(F'12) received the B.S., M.S., and Ph.D. degrees in electrical engineering from the Nanjing Institute of Technology, Nanjing, China, in 1982, 1985,	and 1989, respectively. From 1987 to 1989, he was
	a Lecturer with the Nanjing Institute of Technology.
	Since 1990, he has been with Southeast University,
	first as an Associate Professor and then as a
	Professor. His research interests include mobile communications,
	adaptive signal processing, and artificial
	neural networks with applications to communications
	and biomedical engineering. He contributed
	more than 40 IEEE journal papers and two books in the areas of adaptive signal
	processing and neural networks and their applications to communication
	systems. He was the Premier Foundation Investigator of the China National Science
	Foundation. From 1999 to 2002, he was the Principal Expert of the C3G
	Project, responsible for organizing China's 3G mobile communications research
	and development activities. From 2001 to 2006, he was the Principal Expert of
	the National 863 FuTURE Project. He was the recipient of the Excellent Paper
	Award from the China Institute of Communications in 1987 and the Elite
	Outstanding Young Teacher Award from Southeast University in 1990, 1991,
	and 1993. He is currently the Chairman of the IEEE Nanjing Section. He was
	selected as IEEE Fellow in 2012 for his contributions to the development of
	mobile communications in China.
\end{IEEEbiography}
\begin{IEEEbiography}[{\includegraphics[width=1in,height=1.25in,clip,keepaspectratio]{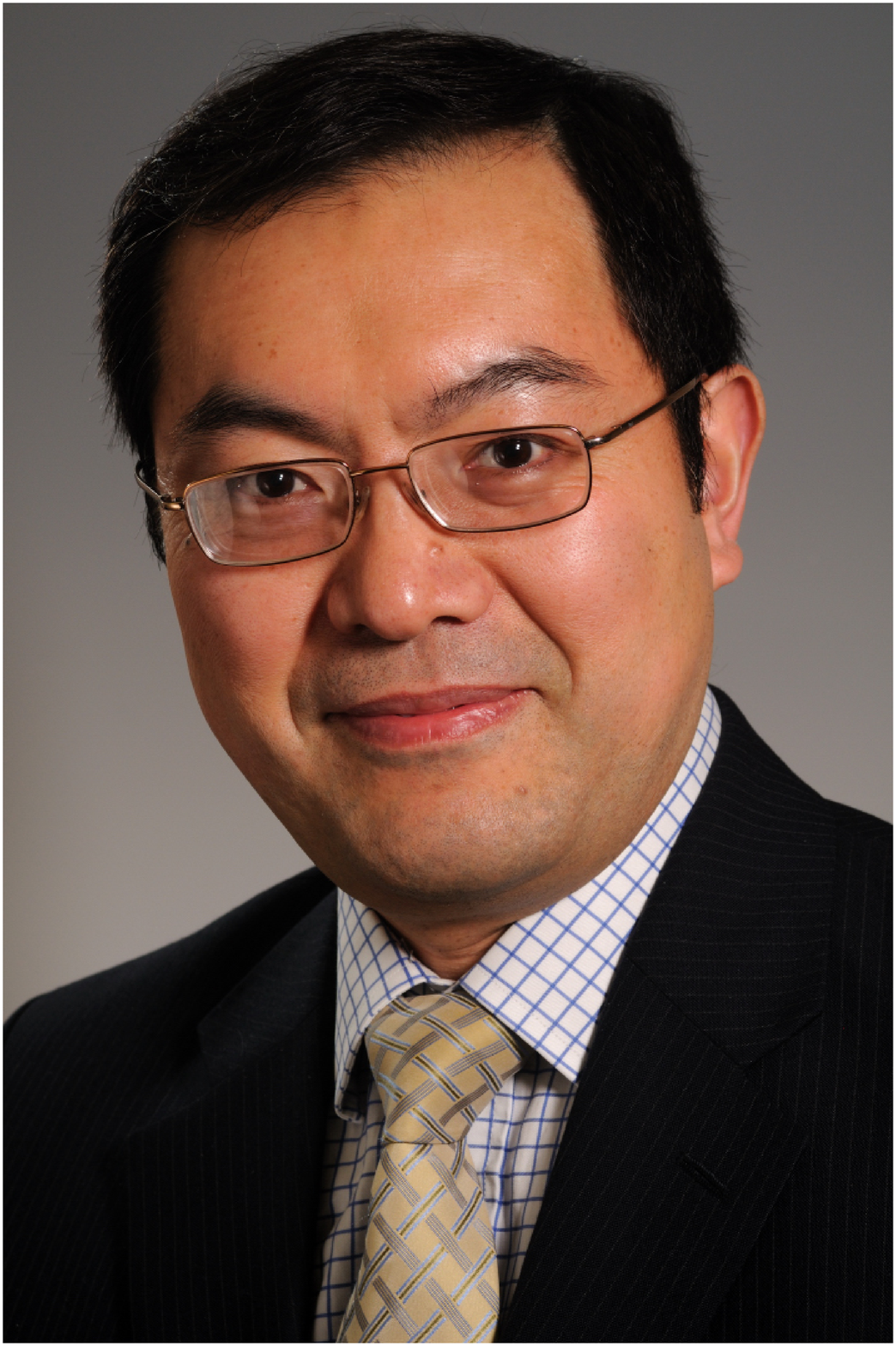}}]{Yang Hao}
(M'00-SM'06-F'13) received the Ph.D.
degree in computational electromagnetics from the
Centre for Communications Research, University of
Bristol, Bristol, U.K., in 1998.

He was a Post-Doctoral Research Fellow with
the School of Electronic, Electrical and Computer
Engineering, University of Birmingham, Birmingham,
U.K. He is currently a Professor of antennas
and electromagnetics with the Antenna Engineering
Group, Queen Mary University of London, London,
U.K. He developed several fully integrated antenna
solutions based on novel artificial materials to reduce mutual RF interference,
weight, cost, and system complexity for security, aerospace, and healthcare,
with leading U.K. industries, novel and emergent gradient index materials to
reduce mass, footprint, and profile of low frequency and broadband antennas,
and also co-developed the first stable active non-Foster metamaterial to
enhance usability through small antenna size, high directivity, and tuneable
operational frequency. He coined the term “body-centric wireless communications,”
i.e., networking among wearable and implantable wireless sensors
on the human body. He was the first to characterize and include the human
body as a communication medium between on-body sensors using surface
and creeping waves. He contributed to the industrial development of the first
wireless sensors for healthcare monitoring. He has authored or co-authored
more than 140 journal papers, and has co-edited and co-authored the books
\textit{Antennas and Radio Propagation for Body-Centric Wireless Communications}
(Artech House, 2006, 2012) and \textit{FDTD Modeling of Metamaterials: Theory
and Applications} (Artech House, 2008), respectively. His current research
interests include computational electromagnetics, microwave metamaterials,
graphene and nanomicrowaves, antennas and radio propagation for bodycentric
wireless networks, active antennas for millimeter/submillimeter applications
and photonic integrated antennas.

Dr. Hao is a Strategic Advisory Board Member of Engineering and
Physical Sciences Research Council, where he is committed to championing
RF/microwave engineering for reshaping the future of UK manufacturing and
electronics.
\end{IEEEbiography}
\end{document}